\documentclass[journal,10pt]{IEEEtran}
\usepackage{epsfig,amsmath,amssymb,epsf,amsthm,scalefnt,multirow}
\usepackage{xcolor}
\usepackage{float}
\usepackage{cite}
\usepackage{psfrag}
\usepackage{graphicx}
\usepackage{subcaption}
\newsavebox{\imagebox}

\newtheorem*{lemma*}{Lemma}

\def\b0{{\pmb{0}}} 

\def\ba{{\mathbf{a}}} \def\bb{{\mathbf{b}}}  
   \def\bh{{\mathbf{h}}}

  \def\bs{{\mathbf{s}}} 
   
\def\by{{\mathbf{y}}} \def\bz{{\mathbf{z}}}  

\def\bA{{\mathbf{A}}} \def\bB{{\mathbf{B}}}  
   \def\bH{{\mathbf{H}}}
   
   \def\bP{{\mathbf{P}}}
  \def\bS{{\mathbf{S}}}

\begin{document}

\title{Radar Imaging Based on IEEE 802.11ad Waveform in V2I Communications}

\author{Geonho Han,~\IEEEmembership{Student Member,~IEEE}, Junil Choi,~\IEEEmembership{Senior Member,~IEEE}, and Robert W. Heath Jr.,~\IEEEmembership{Fellow,~IEEE}
	\thanks{This article was presented in part at the 2020 IEEE Global Communications Conference (GLOBECOM) \cite{HanGLOBECOM:2020} and the 2021 IEEE International Conference on Acoustics, Speech and Signal Processing (ICASSP) \cite{HanICASSP:2021}.
		
		G. Han and J. Choi are with the School of Electrical Engineering, Korea Advanced Institute of Science and Technology, Daejeon 34141, South Korea (e-mail: \{ghhan6, junil\}@kaist.ac.kr).
		
		R. W. Heath Jr. is with the Department of Electrical and Computer Engineering, North Carolina State University, Raleigh NC 27695, USA (e-mail: rwheathjr@ncsu.edu).}}

{}

\maketitle

\begin{abstract}
Since most of vehicular radar systems are already exploiting millimeter-wave (mmWave) spectra, it would become much more feasible to implement a joint radar and communication system by extending communication frequencies into the mmWave band. In this paper, an IEEE 802.11ad waveform-based radar imaging technique is proposed for vehicular settings. A roadside unit (RSU) transmits the IEEE 802.11ad waveform to a vehicle for communications while the RSU also listens to the echoes of transmitted waveform to perform inverse synthetic aperture radar (ISAR) imaging. To obtain high-resolution images of the vehicle, the RSU needs to accurately estimate round-trip delays, Doppler shifts, and velocity of vehicle. The proposed ISAR imaging first estimates the round-trip delays using a good correlation property of Golay complementary sequences in the IEEE 802.11ad preamble. The Doppler shifts are then obtained using least square estimation from the echo signals and refined to compensate phase wrapping caused by phase rotation. The velocity of vehicle is determined using an equation of motion and the estimated Doppler shifts. Simulation results verify that the proposed technique is able to form high-resolution ISAR images from point scatterer models of realistic vehicular settings with different viewpoints. The proposed ISAR imaging technique can be used for various vehicular applications, e.g., traffic condition analyses or advanced collision warning systems.
\end{abstract}

\renewcommand\IEEEkeywordsname{Index Terms}
\begin{IEEEkeywords}
	IEEE 802.11ad, joint radar and communication system, ISAR imaging, vehicular environments.
\end{IEEEkeywords}

\section{Introduction}\label{sec1}
The radar and wireless communication systems have been evolved separately so far to satisfy their performance metrics, e.g., target detection and range/velocity estimation for radar systems \cite{Coluccia:2020,Wang:2021,Baral:2021} and high data rates and good quality of services for wireless communication systems \cite{Cho:2019,Tang:2019,Sohrabi:2021,Wan:2021}. The two systems usually exploit separated frequency bands to avoid interference between them. The communication systems are to make use of millimeter-wave (mmWave) spectra to achieve extremely high data rates. This may lead to a coexistence problem since most of vehicular radar systems have already been occupying the mmWave band. The coexistence of two systems, however, makes it possible to implement a joint radar and communication system (JRCS) using integrated hardware with reduced power consumption and physical space \cite{Mishra:2019,Choi:2016,Liu:2020}.

Since the radar function has done a significant role for real-time collision avoidance and blind spot detection in vehicular settings, we expect the JRCS would be particularly beneficial to vehicle-to-everything (V2X) communications, which will become essential for future vehicles to reduce traffic congestion through optimized driving and to ensure drivers' convenience and safety \cite{Chen:2017,Abboud:2016,Molina:2017,Hobert:2015,Bagheri:2021,Saad:2021}. The communication module of JRCS would be able to obtain side information from the radar function to improve communication performance. For example, radar-aided mmWave communications for V2X can execute efficient beam alignment algorithms by target detection and tracking \cite{Chen:2018,Nuria:2016}. Although legacy vehicular sensors, e.g., camera, radar, or even light detection and ranging (LIDAR), could provide required information for efficient mmWave beam alignment, the joint optimization of wireless communications and sensing in the JRCS can be done by the same manufacture and maximize their potential \cite{Kumari:2017,Dokhanchi:2019}, while it would be quite difficult for the communication module to directly exploit the sensing information from legacy vehicular sensors manufactured by different vendors.

There have been many recent works to realize the JRCS in practice. To add the communication function on radar systems, binary data symbols were embedded on chirp signals in \cite{Saddik:2007}. On the contrary, in \cite{Garmatyuk:2010,Sturm:2011,Keskin:2021}, a popular communication waveform, orthogonal frequency division multiplexing (OFDM), was used to detect targets from the echoes of transmitted signals. The authors in \cite{Chen:2021} proposed code-division OFDM JRCS for machine type communication applications to achieve high spectral efficiency. For OFDM JRCS, intrapulse and intersubcarrier Doppler effects were employed to estimate range and velocity in \cite{Zhang:2020}, and limited feedforward waveform was designed to achieve near-optimal performance trade-off of dual functions in \cite{Keskin:2021_2}. In \cite{Chiriyath:2015,Chiriyath:2017,Bliss:2014}, the performance bounds for data rate and estimation rate as metrics of communications and radar were investigated from an information theoretic viewpoint. The performance trade-off between the Cram\'er-Rao bound (CRB) and distortion minimum mean square error (DMMSE) was considered to design an appropriate JRCS waveform in~\cite{Kumari:2019}. In \cite{Kang:2021}, a radar waveform was designed via maximizing radar performance subject to a sum capacity constraint while \cite{Grossi:2020} tried to maximize the mutual information between input and output symbols for communications by taking radar performance into account.

Beamforming techniques for the JRCS have been also investigated. In \cite{Liu:2018,LiuMU:2018}, optimization problems were formulated to design desired radar beam patterns while simultaneously achieving signal-to-interference-plus-noise ratio (SINR) level for communications in multi-user settings. In \cite{Kumari:2018}, the IEEE 802.11ad waveform was exploited to design beam patterns by leveraging sparsity inherent in the mmWave channels. Radar imaging to recognize objects with high resolution for the JRCS, however, has not been received much interest so far.

The radar imaging was first invented for military use cases \cite{Reamer:1993}, and it has become extremely popular recently due to its versatile applications. The radar imaging can be applied to many fields including image acquisition (map updating), exploration (detecting mineral deposits, oil spills, or small surface movement caused by disasters), or monitoring (climate, agriculture, or tropical forest monitoring) \cite{Inggs:2000} for synthetic aperture radar (SAR), and object classification or deep space imaging of asteroids for inverse synthetic aperture radar (ISAR). The SAR and ISAR are the representative imaging radars with high resolution. The SAR forms an image of a fixed target region by moving a radar in the range direction, where the movement of the radar leads to the synthetic aperture improving the cross-range resolution. The ISAR imaging is for a fixed radar that obtains an image of a moving object by using the range profile and Doppler shifts of the dominant scatterers of the moving target. Contrary to the SAR imaging using radar's movement with planned trajectory, the ISAR imaging suffers from low image quality in general by imperfect knowledge of target motion.

Many previous works have been devoted to improve ISAR image quality through different ways while most of them are based on a pulse compression technique that exploits distinct frequency-varying signals in time to obtain accurate range profiles \cite{Mahafza:2002}. Radar systems usually adopt a linear frequency modulation (LFM) waveform, which is also called as a chirp signal, or a stepped-frequency waveform (SFW) to figure out different look angles of targets \cite{Ozdemir:2012}. In \cite{Zhang:2011}, the sparse SFW (SSFW) by reduced measurement was used to form ISAR images. The potential of a V-style frequency modulation (V-FM) waveform, which is composed of two chirp signals, was investigated for the ISAR imaging in \cite{Gu:2017}. In \cite{Kang:2016}, Tsallis entropy was employed for phase adjustment to reduce computational complexity of conventional phase adjustment method based on Shannon entropy while maintaining ISAR image quality. Compressive sensing with sparse data was exploited for the ISAR imaging in \cite{Kang:2017,Kang2:2017,Zhang:2015}. In \cite{Zhao:2014}, a new autofocus technique based on multitask sparse Bayesian learning was developed. The approximate message-passing network was exploited to perform the ISAR imaging and autofocusing with sparse aperture in \cite{Wei:2021}. All these works, however, are based on pure radar systems without considering any JRCS feature.

Albeit a few, there are some works on the JRCS imaging based on the OFDM waveform in \cite{Garmatyuk:2010,Sturm:2011,Hashempour:2018}. The image quality using OFDM, however, would be degraded in the vehicular environments due to high Doppler shift that could destroy the orthogonality among OFDM subcarriers \cite{Wang:2006}. On the contrary, the IEEE 802.11ad waveform adopted in this paper has several advantages for the ISAR imaging compared to OFDM. First, the IEEE 802.11ad waveform is more robust to the high Doppler shift than OFDM \cite{Duggal:2020}. Second, the ISAR imaging for multiple scatterers located nearby, as in the vehicular environments, requires to use signals with large bandwidth to form high resolution images, and the bandwidth of IEEE 802.11ad waveform is much larger than typical OFDM systems \cite{Heath:2018}. Third, the preamble of IEEE 802.11ad exploits the Golay complementary sequences with a good auto-correlation property, which is beneficial for radar functions. Lastly, the proposed ISAR imaging in this paper only exploits the preamble of IEEE 802.11ad, and obtained sensing parameters could be used for better data communications, which happen after the preamble transmission, through proper mmWave beam alignment.

In this paper, we advance the preliminary JRCS ISAR imaging technique developed in \cite{HanGLOBECOM:2020} to obtain higher resolution images for vehicle-to-infrastructure (V2I) settings. The fundamental radar functions using the IEEE 802.11ad waveform have been developed in \cite{Kumari:2017}, and we extend them to the ISAR imaging. A roadside unit (RSU) for V2I communications is equipped with a radar receiver, which processes the echo signals strongly reflected from a vehicle. The proposed JRCS ISAR imaging technique forms images of the target vehicle from three estimated parameters, i) round-trip delays, ii) Doppler shifts, and iii) vehicular velocity. The preamble embedded on a single-carrier (SC) physical layer (PHY) frame in IEEE 802.11ad has an ideal correlation property suitable for target sensing \cite{Nitsche:2014}, which is exploited for the estimation of round-trip delays. Similar to \cite{HanICASSP:2021}, least square estimation (LSE) of the echo signals recovers the effective radar channels of two specific frames, and we obtain the Doppler shifts using these two estimates, which are then refined to compensate phase wrapping caused by phase rotation. The vehicular velocity is determined from the relationship between an equation of motion and the Doppler shift estimates. In the image formation procedure, we first obtain a pre-image matrix, consisting of range and cross-range profiles, based on the estimated parameters and then form an ISAR image by performing a fast Fourier transform (FFT) of the pre-image matrix along the cross-range direction. We demonstrate via numerical simulations that the proposed JRCS ISAR imaging using the commercialized IEEE 802.11ad standard can achieve high resolution imaging without additional sensors. The high resolution ISAR imaging with precise scaling is possible from the extremely high carrier frequency and wide bandwidth of IEEE 802.11ad waveform. Also, thanks to the accurate Doppler shift estimation method developed in this paper, the ISAR imaging can be performed even for short coherent processing interval (CPI), which is the time that the radar beam scans on a target.

Compared to our previous works in \cite{HanGLOBECOM:2020} and \cite{HanICASSP:2021}, novel contributions in this paper are summarized as follows.

\begin{itemize}
	\item The overall procedure of proposed ISAR imaging technique is clearly shown in Fig. \ref{overall_proc}.
	\item A new cross-range profile is derived to remove ISAR image blurring occurred in \cite{HanGLOBECOM:2020}.
	\item A novel preprocessing technique is developed for the Doppler shift estimation to compensate phase wrapping ambiguity, which was simply ignored in \cite{HanGLOBECOM:2020} and \cite{HanICASSP:2021}, occurring in high velocity environments.
	\item The effectiveness of proposed ISAR imaging technique is verified by extensive simulation results.
	\item The reason of inevitable ISAR image flipping happens in the V2I settings is thoroughly analyzed.
\end{itemize}

Although we do not optimize communication functions with the ISAR imaging technique in this paper, this is an interesting topic that we would like to investigate as our future research topic. The ISAR images obtained using our proposed technique can specify possible positions of a communication transceiver at a vehicle, making it possible to design much more efficient mmWave beam alignment techniques \cite{Nuria:2016}.

The remainder of paper is as follows. In Section II, we first present the preamble structure of IEEE 802.11ad waveform and communication and radar channel models, and then develop transmitted signal and radar received signal models. We propose the delay estimation method employing Golay complementary sequences in Section III and the Doppler shift estimation method in Section IV. Then, we discuss the ISAR image formation procedure in addition to estimating the velocity of vehicle in Section V. In Section VI, we present simulation results with realistic vehicular settings, which show the effectiveness of proposed JRCS ISAR imaging. The conclusion with possible future research topics follows in Section VII.

\begin{figure}
	\centering
	\includegraphics[width=1.0\columnwidth]{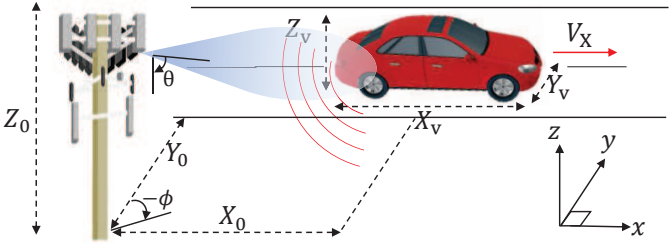}
	\caption{A V2I scenario considered in this paper. The vehicle is represented by its size $(X_{\mathrm{v}},Y_{\mathrm{v}},Z_{\mathrm{v}})$ and velocity $V_{\mathrm{X}}$. The vehicle is located initially from the RSU at $(X_0,Y_0,Z_0)$. The azimuth and elevation angle pair measured from the RSU are represented as $(\phi,\theta)$. It is assumed the distance between RSU and vehicle is relatively larger than the size of vehicle.}\label{V2I}
\end{figure}

\textit{Notation:} $a$, $\ba$, and $\bA$ denote a scalar, vector, and matrix. $|a|$ and $\angle a$ denote the magnitude and angle of $a$. $a[k]$ represents the $k$-th component of $\ba$, and $A[k_1,k_2]$ is the $(k_1,k_2)$-th element of $\bA$. The Euclidean norm of $\ba$ is denoted as $\|\ba\|$. $(\cdot)^{\mathrm{T}}$, $(\cdot)^*$, $(\cdot)^{\mathrm{H}}$, and $(\cdot)^{-1}$ are the transpose, complex conjugate, conjugate transpose, and inverse. The linear convolution of $a(t)$ and $b(t)$ is denoted as $a(t)*b(t)$. The Kronecker product of $\bA$ and $\bB$ is represented as $\bA\otimes\bB$. A complex normal distribution with mean $\mu$ and variance $\sigma^2$ is denoted as $\mathcal{CN}(\mu,\sigma^2)$. $\mathbb{Z}$, $\mathbb{R}$, and $\mathbb{C}$ represent the sets of integer numbers, real numbers, and complex numbers. $\delta[\cdot]$ and $\mathbb{E}[\cdot]$ are the Kronecker delta function and the expectation operation, respectively. $\mathcal{\mathbf{1}}$ denotes the all-ones vector. $\lfloor\cdot\rceil$ and $\lfloor{\cdot\rfloor}$ map its argument into the nearest and the integer part.


\begin{table}
	\centering
	\caption{List of symbols in this paper.}
	\label{t1}
	\begin{tabular}{|c|l|}
		\noalign{\smallskip}\noalign{\smallskip}\hline
		Symbol & Description \\
		\hline
		($\ba_{N}$, $\mathbf{b}_{N}$) & Golay complementary sequences with length $N$ \\
		$M$ & the number of frames in a CPI \\
		$K$ & the number of samples in a frame \\
		$K_{\mathrm{pre}}$  & the number of preamble samples \\
		$K_{\mathrm{c}}$ & Rician $K$-factor \\
		$T_{\mathrm{s}}$ & symbol period \\
		$N_{\mathrm{p}}$ & the number of dominant scatterers \\
		($\phi_p$, $\theta_p$) & azimuth and elevation angles \\
		$\ell_p^m$ & sampled delay \\
		$\nu_p^m$ & Doppler shift \\
		$h_p$ & backscattering coefficient \\
		$\sigma_{\mathrm{nc}}^2$ & variance of noise and clutter \\
		$\check{m}$ & properly selected frame to estimate radar channel \\
		$i$ & frame gap to estimate Doppler shift difference \\
		$\boldsymbol{\Delta}_{\boldsymbol{\nu}}^{m,i}$ & Doppler shift difference vector \\
		$\psi_{p,\mathrm{wrap}}^m$ & residual phase after wrapping in $\hat{\nu}_p^{m}$ \\
		$\psi_{p,\mathrm{wrap}+1}^{m}$ & residual phase after one more wrapping in $\hat{\nu}_p^{m}$ \\
		$\sigma_{\mathrm{wrap}}$ & threshold for preprocessing of phase wrapping \\
		$M_p$ & the number of wrapping \\
		$(\Delta_{\mathrm{r}},\Delta_{\mathrm{cr}})$ & range and cross-range resolution \\
		\hline
	\end{tabular}
\end{table}

\begin{figure*}
	\centering
	\includegraphics[width=1.44\columnwidth]{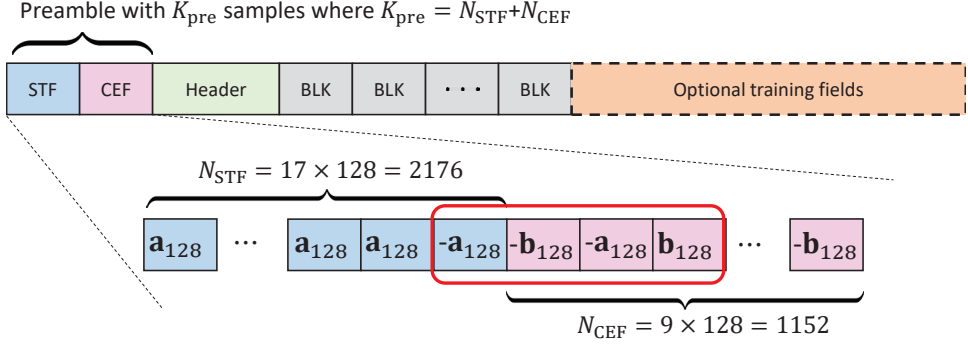}
	\caption{The structure of IEEE 802.11ad SC PHY frame and Golay complementary sequences embedded on an SC PHY preamble, where $\ba_{128}$ and $\mathbf{b}_{128}$ are the Golay complementary bipolar sequences with 128 samples. The segments indicated by the red box are utilized for delay estimation.}\label{Golay}
\end{figure*}

\section{System Model}\label{sec2}
In this paper, we employ the IEEE 802.11ad SC PHY frames to implement the radar imaging functionality. The situation under consideration is that an RSU transmits communication signals to a running vehicle and processes the echo signals at the radar module for the ISAR imaging, as shown in\footnote{While we mainly focus on a single vehicle case in this paper, we briefly discuss how to extend the proposed ISAR imaging technique into the case of multiple vehicles in Remark 3 in Section V.} Fig. \ref{V2I}. We first look into the structure of preamble embedded in the IEEE 802.11ad waveform. We then describe V2I communication channel and two-way multi-target radar channel models. Finally, communication transmitted signal and radar received signal models based on the IEEE 802.11ad waveform are developed. Some important symbols used throughout the paper are listed in TABLE I.

\subsection{Preamble structure in the IEEE 802.11ad waveform}
The SC PHY frame of IEEE 802.11ad consists of a preamble, which contains short training field (STF) and channel estimation field (CEF), a header, data blocks (BLKs), and optional training fields, as shown in Fig. \ref{Golay}. The STF is used for frame synchronization and frequency offset estimation, and the CEF is for channel estimation. Fig. \ref{Golay} shows both of them are composed of Golay complementary sequences $\ba_{128}$ and $\bb_{128}$, where $K_{\mathrm{pre}}$ is the number of preamble samples. We will focus on the preamble of IEEE 802.11ad waveform that can be exploited to estimate parameters required for the ISAR imaging from echo signals.

\subsection{Communication and radar channel models}
We assume the communication channel follows the Rician fading since the radar module at the RSU requires to have the line-of-sight (LOS) path to detect targets of interest. For the RSU to perform the ISAR imaging of a vehicle, the vehicle is assumed to consist of multiple dominant scatterers, leading to a multi-target model. In this paper, we focus on the scenario in Fig. 1 with a single object typically considered in the ISAR imaging \cite{Ozdemir:2012}. It would be highly likely that the communication transceiver on the vehicle becomes one of the multiple dominant scatterers since the transceiver would be judiciously deployed on the vehicle to experience the LOS path for communications with a high probability \cite{Nuria:2016}.

We consider a V2I multiple-input multiple-output (MIMO) system with $N_{\mathrm{vRX}}$ receive (RX) antennas at the vehicle and $N_{\mathrm{TX}}$ transmit (TX) antennas at the RSU for communications. The RSU also has $N_{\mathrm{RX}}$ RX antennas for the radar module. While they can be different, we assume $N_{\mathrm{TX}}=N_{\mathrm{RX}}$ for simplicity. Both the RSU and vehicle are assumed to have uniform planar arrays (UPAs), which consist of antennas along the $x$-axis (horizontal axis) and $y$-axis (vertical axis), resulting in $N_{\mathrm{TX}}=N^x_{\mathrm{TX}}\times N^y_{\mathrm{TX}}=N_{\mathrm{RX}}=N^x_{\mathrm{RX}}\times N^y_{\mathrm{RX}}$ and $N_{\mathrm{vRX}}=N^x_{\mathrm{vRX}}\times N^y_{\mathrm{vRX}}$.

The communication channel of the $m$-th frame in a CPI is represented as
\begin{equation}\label{eq1}
\mathbf{H}_{\mathrm{com}}[m]= \sqrt{\frac{K_{\mathrm{c}}}{K_{\mathrm{c}}+1}}\mathbf{H}_{\mathrm{LOS}}[m]+\sqrt{\frac{1}{K_{\mathrm{c}}+1}}\mathbf{H}_{\mathrm{NLOS}}[m],
\end{equation}
where $K_{\mathrm{c}}$ is the Rician $K$-factor that measures how strong the LOS channel is \cite{Mukherjee:2017}. Using the superscript $(\cdot)^m$ to indicate the $m$-th frame, the LOS channel is represented as
\begin{align}\label{eq2}
\mathbf{H}_{\mathrm{LOS}}&[m]=\notag\\
&\alpha_0^me^{j2\pi\nu_{\mathrm{com}}^mmKT_{\mathrm{s}}}\mathbf{a}_{\mathrm{vRX}}(\phi_{\mathrm{com}}^m,\theta_{\mathrm{com}}^m)\mathbf{a}^{\mathrm{H}}_{\mathrm{TX}}(\phi_{\mathrm{com}}^m,-\theta_{\mathrm{com}}^m),
\end{align}
where $\alpha_0^m$ is the complex path gain of LOS path distributed as $\mathcal{CN}(0,1)$, $\nu_{\mathrm{com}}^m$ represents the Doppler shift experienced at the vehicle, i.e., the communication receiver, $K$ is the number of samples in a frame, $T_{\mathrm{s}}$ represents the symbol period, and $\theta_{\mathrm{com}}^m$ and $\phi_{\mathrm{com}}^m$ denote the elevation and azimuth angle of arrivals (AoAs) at the vehicle. In (\ref{eq2}), $\mathbf{a}_{\mathrm{TX}}(\phi,\theta)$ is the TX array steering vector at the RSU, and $\mathbf{a}_{\mathrm{vRX}}(\phi,\theta)$ is the RX array steering vector at the vehicle. The sign of angle $\theta_{\mathrm{com}}^m$ at the RSU becomes the opposite to that at the vehicle due to the relative height gap between them, as described in Fig. \ref{V2I}. The non-line-of-sight (NLOS) channel is given as
\begin{align}\label{eq3}
\bH_{\mathrm{NLOS}}[m]=&\frac{1}{\sqrt{N_{\mathrm{NLOS}}}}\sum_{u=1}^{N_{\mathrm{NLOS}}}\alpha_u^m e^{j2\pi \nu_{\mathrm{com}}^mmKT_{\mathrm{s}}}\notag\\
&\,\,\times\ba_{\mathrm{vRX}}(\phi_{\mathrm{vRX},u}^m,\theta_{\mathrm{vRX},u}^m \ba^{\mathrm{H}}_{\mathrm{TX}}(\phi_{\mathrm{TX},u}^m,\theta_{\mathrm{TX},u}^m),
\end{align}
where $N_{\mathrm{NLOS}}$ is the number of NLOS paths, $\alpha_u^m$ is the complex path gain distributed as $\mathcal{CN}(0,1)$, and $(\phi_{\mathrm{vRX},u}^m,\theta_{\mathrm{vRX},u}^m)$ and $(\phi_{\mathrm{TX},u}^m,\theta_{\mathrm{TX},u}^m)$ denote the azimuth and elevation angle pairs at the vehicle and RSU of the $u$-th NLOS path.

It would be possible for the RSU to perform full-duplex communication and radar operations if the TX antennas (for radar and communications) and RX antennas (for radar) are closely separated with analog/digital self-interference cancellation (SIC) techniques \cite{Hong:2014}. The radar channel model is then represented as
\begin{align}\label{eq4}
\mathbf{H}_{\mathrm{rad}}(t)=&\sum_{p=0}^{N_{\mathrm{p}}-1}\sqrt{G_p(t)}\beta_p(t) e^{j2\pi t\nu_p(t)}e^{-j2\pi f_{\mathrm{c}}\tau_p(t)}\notag\\
&\qquad\quad\,\,\times\mathbf{a}^*_{\mathrm{RX}}(\phi_p(t),\theta_p(t))\mathbf{a}^{\mathrm{H}}_{\mathrm{TX}}(\phi_p(t),\theta_p(t)),
\end{align}
where $N_{\mathrm{p}}$ is the number of dominant scatterers. Several parameters relevant to the $p$-th dominant scatterer in (\ref{eq4}) are denoted as follows: $\beta_p(t)\sim\mathcal{CN}(0,1)$ is the small-scale complex channel gain, $\phi_p(t)$ and $\theta_p(t)$ are the azimuth and elevation AoAs, and the large-scale channel gain $G_p(t)=\lambda^2\sigma_p^{\mathrm{RCS}}/(64\pi^3r_p^4(t))$ includes the effects of path-loss and radar cross section (RCS), which is a measure of how detectable a target is, where $\sigma_p^{\mathrm{RCS}}$ and $r_p(t)$ are the RCS and Euclidean distance, and $\lambda$ represents the wavelength corresponding to the carrier frequency $f_{\mathrm{c}}$. The Doppler shift is $\nu_p(t)=2v_p(t)/\lambda$ with the relative velocity $v_p(t)=V_{\mathrm{X}}\mathrm{sin}(\phi_p(t))$ from the viewpoint of RSU where $V_{\mathrm{X}}$ is the actual velocity of vehicle. The round-trip delay is $\tau_p(t)=2r_p(t)/c$ with the speed of light $c$. In (\ref{eq4}), the conjugate on the RX array steering vector at the RSU, $\ba_{\mathrm{RX}}^*(\phi_p(t),\theta_p(t))$, is due to co-located TX and RX antennas from the $p$-th dominant scatterer \cite{Jian:2007}. Note that the sizes of $\ba_{\mathrm{RX}}(\phi_p(t),\theta_p(t))$ and $\ba_{\mathrm{TX}}(\phi_p(t),\theta_p(t))$ are the same since we assume $N_{\mathrm{TX}}=N_{\mathrm{RX}}$.

We define spatial frequencies
\begin{equation}\label{eq5}
\varphi_x=\frac{2\pi d_x\mathrm{cos}(\theta)\mathrm{sin}(\phi)}{\lambda},\quad\varphi_y=\frac{2\pi d_y\mathrm{sin}(\theta)}{\lambda},
\end{equation}
where $d_x$ and $d_y$ denote the antenna spacing for the horizontal and vertical directions. By adopting the half wavelength spacing for both horizontal and vertical antennas, the array steering vector is then represented as
\begin{equation}\label{eq6}
\ba_{\mathrm{AX}}(\phi,\theta)=\begin{bmatrix}
1\\
e^{j\varphi_x}\\
\vdots\\
e^{j(N^x_{\mathrm{AX}}-1)\varphi_x}
\end{bmatrix}\otimes\begin{bmatrix}
1\\
e^{j\varphi_y}\\
\vdots\\
e^{j(N^y_{\mathrm{AX}}-1)\varphi_y}
\end{bmatrix},
\end{equation}
where AX $\in\left\{\mathrm{TX},\mathrm{RX},\mathrm{vRX}\right\}$.

\subsection{Transmit and received signal models}
To \textit{see} a vehicle, consisting of multiple dominant scatterers, with mobility, the RSU must listen to the echoes of multiple transmitted frames. A transmitted signal at the RSU employing the IEEE 802.11ad waveform consists of $M$ frames for one CPI, where one frame contains $K$ samples. The continuous time representation of TX baseband signal is
\begin{equation}\label{eq7}
x(t)=\sqrt{\mathcal{E}_s}\sum_{n=-\infty}^{\infty}s[n]g_{\mathrm{TX}}(t-nT_{\mathrm{s}}),
\end{equation}
where $s[n]$ denotes the normalized TX symbol of IEEE 802.11ad waveform, $\mathcal{E}_s$ is the symbol energy, and $g_{\mathrm{TX}}(t)$ is the TX pulse shaping filter.

\begin{figure}
	\centering
	\includegraphics[width=0.898\columnwidth]{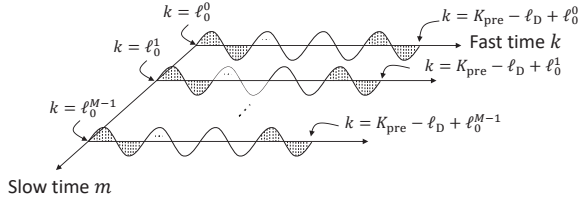}
	\caption{Fast and slow time representation of radar received signals.}\label{fast_slow_time}
\end{figure}

\begin{figure*}
	\centering
	\includegraphics[width=1.32\columnwidth]{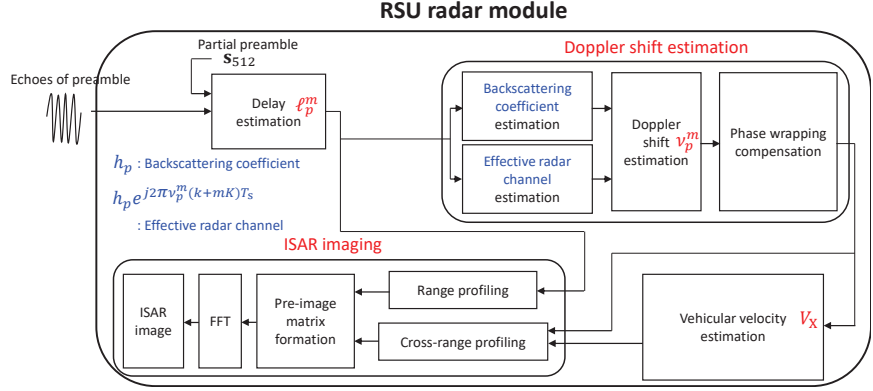}
	\caption{Overall procedure of proposed ISAR imaging.}\label{overall_proc}
\end{figure*}

To implement radar imaging operation, the RSU needs to handle the echo signals at the receiving module.\footnote{While IEEE 802.11ad operates with time division duplexing (TDD) mode \cite{Chen:2019}, the uplink signals from the vehicle would not cause any interference to the echoes of preamble since the echo signals are received much earlier than the uplink transmission.} Assuming the multi-target model for a vehicle, the radar received signal passed through a matched filter $g_{\mathrm{RX}}(t)$ with the same roll-off factor as $g_{\mathrm{TX}}(t)$ becomes
\begin{equation}\label{eq8}
y(t)=\sum_{p=0}^{N_{\mathrm{p}}-1}\sqrt{\mathcal{E}_s}h_p(t) x_g (t-\tau_p(t))e^{j2\pi  t\nu_p(t)}+z(t)
\end{equation}
with
\begin{align}\label{eq9}
h_p(t)=\sqrt{G_p(t)}\beta_p(t) \mathbf{f}^{\mathrm{H}}_{\mathrm{RX,rad}}&\mathbf{a}^*_{\mathrm{RX}}(\phi_p(t),\theta_p(t))\notag\\
&\quad\times \mathbf{a}^{\mathrm{H}}_{\mathrm{TX}}(\phi_p(t),\theta_p(t))\mathbf{f}_{\mathrm{TX}},
\end{align}
where $x_g(t)=\sum_{n=-\infty}^{\infty} s[n] g(t-nT_{\mathrm{s}})$ with $g(t)=g_{\mathrm{TX}}(t)*g_{\mathrm{RX}}(t)$. The TX pulse shaping filter and RX matched filter meet the Nyquist criterion, i.e., $g(nT_{\mathrm{s}})=\delta[n]$. The term $z(t)\sim\mathcal{CN}(0,\sigma_{\mathrm{nc}}^2)$ incorporates the effect of noise and clutter reflected from undesirable objects\footnote{Because of the beamforming with narrow beamwidth in IEEE 802.11ad, there would be no significant clutter by effective scatterers except the desired targets \cite{Currie:1987}. Note that the effect of insignificant clutter, e.g., the echo reflected from a road, is typically assumed as noise \cite{Kumari:2017,Chiriyath:2015_clutter}. Thus, the effect of insignificant clutter is included in $z(t)$ of (\ref{eq8}) having the variance $\sigma_{\mathrm{nc}}^2=N_0W+P_{\mathrm{c}}$, where $N_0$ denotes the noise spectral density, $W$ is the bandwidth, and $P_{\mathrm{c}}$ represents the average clutter power \cite{skolnik:1980}.} \cite{Shnidman:1999,Mahafza:2002}. The backscattering coefficient $h_p(t)$ in (\ref{eq9}) determines the echo signal strength from the $p$-th dominant scatterer, where $\mathbf{f}_{\mathrm{RX,rad}}$ and $\mathbf{f}_{\mathrm{TX}}$ denote the RX combiner for the radar module and the TX beamformer for the communication module. Due to the short distance that the target vehicle moves for sufficiently short CPI, the variation of time-related parameters in $h_p(t)$ is almost constant and insignificant for the ISAR imaging. Therefore, the backscattering coefficients for all dominant scatterers are assumed to be constant in our model, i.e., $h_p\approx\sqrt{G_p}\beta_p \mathbf{f}^{\mathrm{H}}_{\mathrm{RX,rad}}\mathbf{a}^*_{\mathrm{RX}}(\phi_p,\theta_p)\mathbf{a}^{\mathrm{H}}_{\mathrm{TX}}(\phi_p,\theta_p)\mathbf{f}_{\mathrm{TX}}$.

In the JRCS setting, the communication module at the RSU would be able to inform $\mathbf{f}_{\mathrm{RX,rad}}$ to the radar module. To be specific, the TX beamformer at the RSU and RX combiner at the vehicle for communications are obtained through the beam alignment process at $m=0$ as \cite{Zhou:2012}
\begin{equation}\label{eq10}
(\hat{\mathbf{f}}_{\mathrm{TX}},\hat{\mathbf{f}}_{\mathrm{RX}})=\underset{\mathbf{f}_{\mathrm{TX}}\in\mathcal{W}_{\mathrm{T}},\mathbf{f}_{\mathrm{RX}}\in\mathcal{W}_{\mathrm{R}}}{\mathrm{argmax}} |\mathbf{f}_{\mathrm{RX}}^{\mathrm{H}}\bH_{\mathrm{com}}[0]\mathbf{f}_{\mathrm{TX}}|^2,
\end{equation}
where $\mathbf{f}_{\mathrm{RX}}$ is the RX beamformer at the target vehicle, $\mathcal{W}_{\mathrm{T}}$ and $\mathcal{W}_{\mathrm{R}}$ are the codebooks consisting of the $N_{\mathrm{TX}}\times 1$ and $N_{\mathrm{RX}}\times 1$ discrete Fourier transform (DFT) vectors, respectively.\footnote{We assume the sizes of $\mathcal{W}_{\mathrm{T}}$ and $\mathcal{W}_{\mathrm{R}}$ are $N_{\mathrm{TX}}$ and $N_{\mathrm{RX}}$, respectively, while the codebook size is a system parameter that can be further optimized.} The radar module at the RSU then can set $\mathbf{f}_{\mathrm{RX,rad}}=\mathbf{f}_{\mathrm{TX}}^*$ to maximize radar beamforming gain $|\mathbf{f}_{\mathrm{RX,rad}}^{\mathrm{H}}\bH_{\mathrm{rad}}(t)\mathbf{f}_{\mathrm{TX}}|^2$ since the transmit signal and its echo signal experience the reciprocal channels just like in the TDD systems. Note that the RX beamformer $\mathbf{f}_{\mathrm{RX}}$ in (\ref{eq10}) is selected at the target vehicle, and $\mathbf{f}_{\mathrm{RX}}\neq \mathbf{f}_{\mathrm{TX}}$ in general even for the TDD systems. We assume $\mathbf{f}_{\mathrm{TX}}$ and $\mathbf{f}_{\mathrm{RX,rad}}$ are fixed for one CPI considering sufficiently short CPI.

Before deriving the discrete-time radar received signal model, we first show the echoes of preamble of interest in the fast and slow time domains in Fig. \ref{fast_slow_time}, where $k$ and $m$ denote the sample and frame indices. In Fig. \ref{fast_slow_time}, $\ell_0^m$ is the first sampled delay caused by transmitted samples in the $m$-th frame, where the sampled delays are positive integers within each frame after sampling the echoes at the RSU. To avoid the interference from the echoes of data streams, with a design parameter $\ell_{\mathrm{D}}$ that is a positive integer, we only consider the echo sample index from $\ell_0^m$ to $K_{\mathrm{pre}}-\ell_{\mathrm{D}}+\ell_0^m$ in each frame for the proposed ISAR imaging since the first echo sample index due to the data streams is $k=K_{\mathrm{pre}}+\ell_0^m$. Although it is enough to have $\ell_{\mathrm{D}}=1$ to avoid the interference ideally, in practice, it may be better to have $\ell_{\mathrm{D}}$ larger than one to make the system robust to possible interference from the data streams.

The discrete-time sampled representation of radar received signal in (\ref{eq8}) is then given as
\begin{align}\label{eq11}
y(nT_{\mathrm{s}})=&\sum_{p=0}^{N_{\mathrm{p}}-1}\sqrt{\mathcal{E}_s}h_p x_g(nT_{\mathrm{s}}-\tau_p(nT_{\mathrm{s}}))e^{j2\pi nT_{\mathrm{s}}\nu_p(nT_{\mathrm{s}})}\notag\\
&+z(nT_{\mathrm{s}})\notag\\
=&\sum_{p=0}^{N_{\mathrm{p}}-1}\sqrt{\mathcal{E}_s}h_p x_g((k-\ell_p^m)T_{\mathrm{s}})e^{j2\pi\nu_{p}^m(k+mK)T_{\mathrm{s}}}\notag\\
&+z((k+mK)T_{\mathrm{s}}),
\end{align}
where $n=k+mK$ for $k=\ell_0^m,\ell_0^m+1,\cdots,K_{\mathrm{pre}}-\ell_{\mathrm{D}}+\ell_0^m$ and $m=0,1,\cdots,M-1$. The Doppler shift of $p$-th dominant scatterer after sampling can be represented as
\begin{equation}\label{n1}
\nu_p(nT_s)=\nu_p^m+\nu_p(kT_s)\approx\nu_p^m.
\end{equation}
In (\ref{n1}), we only consider the Doppler shift variation with frames $\nu_p^m$ due to the negligible change of $\nu_p(kT_s)$ during the sample period. The round-trip delay due to the $p$-th dominant scatterer after sampling is given by
\begin{equation}\label{eq12}
\tau_p(nT_{\mathrm{s}})=mKT_{\mathrm{s}}+\ell_p^mT_{\mathrm{s}}+\tau_p^{\mathrm{f}}(nT_{\mathrm{s}}),
\end{equation}
where $\ell_p^m$ is the sampled delay within the $m$-th frame that would be recovered from the received signal, and $\tau_p^{\mathrm{f}}(t)$ is the unknown fractional part of delay after sampling, which is negligible when the sampling period $T_{\mathrm{s}}$ is extremely small as in the IEEE 802.11ad waveform. The discrete-time representation of the radar received signal can be written as
\begin{equation}\label{eq13}
y[m,k]=\sum_{p=0}^{N_{\mathrm{p}}-1}\sqrt{\mathcal{E}_s}h_p e^{j2\pi \nu_{p}^m (k+mK)T_{\mathrm{s}}} s[k-\ell_p^m]+z[m,k],
\end{equation}
where
\begin{align}\label{eq14}
s[k-\ell_p^m]&=\sum_{n'=-\infty}^{\infty}s[n']g((k-\ell_p^m-n')T_{\mathrm{s}})\notag\\
&=x_g((k-\ell_p^m)T_{\mathrm{s}})
\end{align}
with $g(t)$ satisfying the Nyquist criterion.


To form the ISAR image, the RSU needs to estimate the sampled delays $\ell_p^m$ and Doppler shifts $\nu_{p}^m$ for the multiple scatterers strongly reflected from the vehicle, and the vehicular velocity $V_{\mathrm{X}}$. The overall procedure of proposed ISAR imaging is described in Fig. \ref{overall_proc} where detailed methods are developed in the following sections.

\section{Delay Estimation}
The IEEE 802.11ad waveform exploits Golay complementary sequences, which consist of bipolar sequences with a good correlation property, for the preamble \cite{Parker:2003}. The correlation property of Golay complementary bipolar sequences with $N$ samples, which are $\mathbf{a}_N$ and $\mathbf{b}_N$, is given as
\begin{equation}\label{eq15}
R_{{\mathbf{a}}_N}[k]+R_{{\mathbf{b}}_N}[k]=2N\delta[k],
\end{equation}
where
\begin{equation}\label{eq16}
R_{\mathbf{c}}[k]=\sum_{q=0}^{N-k-1}c[q]c[q+k],
\end{equation}
for an arbitrary vector $\mathbf{c}$ of length $N$. The correlation property in (\ref{eq15}) will be used to estimate the delays from strongly reflected echo signals similar to \cite{Kumari:2017,Muns:2019} but with a different part of preamble to exploit a better correlation property. Since the RSU knows its transmitted symbols, the symbols other than the preamble, e.g., data symbols, could be used for the auto-correlation as well. The data symbols, however, may not have a good correlation property in general, which would cause incorrect delay estimation.

To estimate the sampled delays $\ell_p^m$, we first derive a correlation function with the auto-correlation property using certain segments of preamble. Then, the sampled delays are obtained by thresholding with the upper bound of the noisy part of correlation function.

\subsection{Correlation function with auto-correlation property}
We focus on several segments of length 512 samples, which correspond to the endpiece of STF and the forepart of CEF embedded on the SC PHY preamble, as marked with the red box in Fig. \ref{Golay}. These segments have the ideal auto-correlation property up to 128 and 64 samples backward and forward, respectively. Therefore, their echoes are interfered with neither the echoes of data streams (as discussed in Section II) nor the echoes of preamble close to these 512 samples due to the ideal auto-correlation property. Since the segments are starting from the 2049-th sample, we explicitly consider this offset in the remaining discussions.

The set of sampled delays for the $m$-th frame is extracted from the echoes of preamble by using a correlation function, just like the matched filtering typically used in many radar systems to obtain range profile, defined as
\begin{align}\label{eq17}
\tilde{R}_{\bs_{512}\tilde{\by}_m}[\ell]&=\sum_{k_N=0}^{511} {s_{512}}[k_N]\tilde{y}_m^*[\ell+k_N]\notag\\
&=\sum_{k_N=0}^{511} {s_{512}}[k_N]y^*[m,\ell+k_N+2048]\notag\\
&=\sum_{k_N=0}^{511}\sum_{p=0}^{N_{\mathrm{p}}-1} \sqrt{\mathcal{E}_s}h^*_pe^{-j2\pi\nu_{p}^m(\ell+k_N+2048+mK)T_{\mathrm{s}}}\notag \\
&\,\,\,\,\times s_{512}[k_N]s[k_N+2048+\ell-\ell_p^m]+\tilde{\bz}_m^{\mathrm{H}}\bs_{512},
\end{align}
where $\tilde{y}_m[k]=y[m,k+2048]$, $\tilde{z}_m[k]=z[m,k+2048]$, and $\bs_{512}=\begin{bmatrix}\mathbf{-a}^{\mathrm{T}}_{128}\,\mathbf{-b}^{\mathrm{T}}_{128}\,\mathbf{-a}^{\mathrm{T}}_{128}\,\mathbf{b}^{\mathrm{T}}_{128}\end{bmatrix}^{\mathrm{T}}$. Note that the $k$-th elements of $\tilde{\by}_m$, $\bs_{512}$, and $\tilde{\bz}_m$ are $\tilde{y}_m[k]$, $s_{512}[k]$, and $\tilde{z}_m[k]$, respectively. Although it is possible to use more Golay complementary sequences for the correlation function to have a larger peak and detect the dominant scatterers experiencing small backscattering coefficient $h_p$, this breaks the ideal auto-correlation property and makes the possible range of delay estimation narrower since the sample index in $y[m,\ell+k_N+2048]$ is constrained as $\ell+k_N+2048\leq K_{\mathrm{pre}}-\ell_{\mathrm{D}}+\ell_0^m$, i.e., the search range $\ell$ decreases with the increase of maximum value of $k_N$ that corresponds to the length of segments used for the correlation function. This is the reason we only exploited the segments marked by the red box in Fig. \ref{Golay} for delay estimation.

\subsection{Sampled delay extraction}
The radar module at the RSU is able to estimate the sampled delay from the dominant scatterer as
\begin{equation}\label{eq18}
\hat{\ell}_{\mathrm{peak}}^m=\mathrm{argmax}_{\ell}\,\,|\tilde{R}_{\bs_{512}\tilde{\by}_m}[\ell]|.
\end{equation}
By searching $\ell$, which results in $|\tilde{R}_{\bs_{512}\tilde{\by}_m}[\ell]|$ greater than a certain threshold, both backward and forward from $\hat{\ell}_{\mathrm{peak}}^m$, the RSU can estimate the remaining sampled delays from other dominant scatterers on the vehicle.\footnote{It is possible to just select all the sampled delays that result in $|\tilde{R}_{\bs_{512}\tilde{\by}_m}[\ell]|$ larger than a threshold. This may include, however, some echoes from other parts of preamble located far from the segments of interest, leading to incorrect estimated sampled delays.} Note that the RSU can only search several sampled delays from $\hat{\ell}_{\mathrm{peak}}^m$ since the size of vehicle is typically less than a few meters. From (\ref{eq17}), we use the upper bound of $\tilde{\mathbf{z}}_m^{\mathrm{H}}\bs_{512}$ for the threshold, which is easily deduced by the Cauchy-Schwarz inequality as
\begin{equation}\label{eq19}
\tilde{\bz}_m^{\mathrm{H}}\bs_{512}\le\|\bs_{512}\|\cdot\|\tilde{\bz}_m\|=512\cdot\sigma_{\mathrm{nc}}=\sigma_{\mathrm{th}}.
\end{equation}
We assume all the estimated sampled delays are mutually distinct at $m=0$ due to the wide bandwidth of IEEE 802.11ad waveform. The set of sampled delays for $m=0$ is then given as
\begin{equation}\label{eq20}
\mathcal{L}_0=\left\{\hat{\ell}_0^0, \hat{\ell}_1^0,\cdots, \hat{\ell}_{\mathrm{peak}}^0,\cdots, \hat{\ell}_{\hat{N}_{\mathrm{p}}-2}^0, \hat{\ell}_{\hat{N}_{\mathrm{p}}-1}^0\right\},
\end{equation}
where $\hat{N}_{\mathrm{p}}$ is the number of estimated dominant scatterers on the vehicle, and $\hat{\ell}_{x_1}^0<\hat{\ell}_{x_2}^0$ for $x_1<x_2$. Even though typical radar systems used for target detection need to detect all the targets of interest, this is not the case for the ISAR imaging. If the number of estimated dominant scatterers is less than that of true dominant scatterers, the RSU still can obtain the ISAR images having the same shape of the target with some blank holes. For the opposite case, there will be some additional points inside or outside of the ISAR images, which can be easily compensated with advanced image processing techniques. Therefore, as long as the number of estimated dominant scatterers is close to the true value, which can be done by the threshold in (\ref{eq19}), there will be no issue for obtaining proper ISAR images.

Depending on the sampled fractional delay $\tau_p^{\mathrm{f}}(nT_{\mathrm{s}})$ in (\ref{eq12}), it is possible that $\mathcal{L}_m$ varies with frames, and several sampled delays in $\mathcal{L}_m$ may overlap as time evolves. This issue should be resolved to obtain high resolution ISAR images, which is discussed in the next section.

\section{Doppler Shift Estimation}
There could be many different ways to estimate the Doppler shifts of dominant scatterers. In this paper, we first obtain i) the backscattering coefficients $h_p$ and ii) the effective radar channel estimates $h_pe^{j2\pi\nu_p^m(k+mK)T_{\mathrm{s}}}$ from the $0$-th and $m$-th frames, respectively, and then extract iii) the Doppler shifts $\nu_p^m$ from the estimated parameters. Finally, we obtain accurate Doppler shifts for all frames after compensating the phase wrapping effect.

\subsection{Backscattering coefficient estimation}
To estimate $h_p$ first, we take the $0$-th frame of the received signal into account as
\begin{align}\label{eq21}
y[0,k]&=\sum_{p=0}^{N_{\mathrm{p}}-1}\sqrt{\mathcal{E}_s}h_p e^{j2\pi \nu_p^0kT_{\mathrm{s}}}s[k-\ell_p^0]+z[0,k]\notag\\
&\stackrel{(a)}{\approx}\sum_{p=0}^{\hat{N}_{\mathrm{p}}-1}\sqrt{\mathcal{E}_s}s[k-\hat{\ell}_p^0]h_p+z[0,k],
\end{align}
for $k=\hat{\ell}_{0}^0,\hat{\ell}_{0}^0+1,\cdots,K_{\mathrm{pre}}-\ell_{\mathrm{D}}+\hat{\ell}_0^0$, where $(a)$ comes from the approximation $e^{j2\pi\nu_p^0kT_{\mathrm{s}}}\approx 1$, which holds due to the extremely short symbol period $T_{\mathrm{s}}$ of the IEEE 802.11ad waveform. Note that $\hat{\ell}_0^0$ represents the first sampled delay of echo signal the radar module receives at the $0$-th frame. By concatenating $K_{\mathrm{pre}}-\ell_{\mathrm{D}}+1$ received samples, i.e., all the echo signals of preamble, we have
\begin{equation}\label{eq22}
\by_0=\sqrt{\mathcal{E}_s}\bS_0\bh_0+\bz_0,
\end{equation}
where
\begin{align}\label{eq23}
\by_0&=[y[0,\hat{\ell}_{0}^0],y[0,\hat{\ell}_{0}^0+1],\cdots,y[0,K_{\mathrm{pre}}-\ell_{\mathrm{D}}+\hat{\ell}_0^0]]^{\mathrm{T}},\notag\\
\bh_0&=[h_0,h_1,\cdots,h_{\hat{N}_{\mathrm{p}}-1}]^{\mathrm{T}},\notag\\
\bz_0&=[z[0,\hat{\ell}_{0}^0],z[0,\hat{\ell}_{0}^0+1],\cdots,z[0,K_{\mathrm{pre}}-\ell_{\mathrm{D}}+\hat{\ell}_0^0]]^{\mathrm{T}},
\end{align}
and the ($x_1,x_2$)-th element of $\bS_0\in\mathbb{Z}^{(K_{\mathrm{pre}}-\ell_{\mathrm{D}}+1)\times\hat{N}_{\mathrm{p}}}
$ is $s[\hat{\ell}_{0}^0+x_1-1-\hat{\ell}_{x_2-1}^0]$ if $0\le\hat{\ell}_{0}^0+x_1-1-\hat{\ell}_{x_2-1}^0\le K_{\mathrm{pre}}-1$ while all other elements of $\bS_0$ are zero. Then, the estimate of $\bh_0$ is obtained by LSE as
\begin{equation}\label{eq24}
\hat{\bh}_0=\frac{(\bS_0^\mathrm{H}\bS_0)^{-1}\bS_0^\mathrm{H}\by_0}{\sqrt{\mathcal{E}_s}}.
\end{equation}
Even though it is possible to use not just the $0$-th frame but also its nearby frames to estimate $h_p$, the numerical results in Section VI show that the backscattering coefficients are estimated well from the received signals of the $0$-th frame with the long enough length of $K_{\mathrm{pre}}-\ell_D$. Therefore, the backscattering coefficient estimation using more frames may just slow down the signal processing without any benefit.

\subsection{Effective radar channel estimation}
Now, we need to have the effective radar channel estimates of the $m$-th frame. While any frame (except $m=0$ that is already used) is possible, it is better to take a small frame index not to incur possible overlapped delays as discussed at the end of Section III. With too small $m$, however, the approximation error in (\ref{eq21}) would not be mitigated. In Section IV-C, we briefly discuss how to select a proper frame index $m$ to avoid the approximation error. Also, we numerically verify this in Section VI and show that the range of proper value of $m$ is large, which makes the proposed technique more practical.

By denoting a properly selected frame index as $\check{m}$, the radar channels of the $\check{m}$-th frame can be estimated similarly with the case of the $0$-th frame, but the approximation, $e^{j2\pi\nu_{p}^{\check{m}} (k+\check{m}K)T_{\mathrm{s}}}\approx1$, may not hold for $\check{m}>0$. Instead, using the fact $k\ll \check{m}K$, we can approximate (\ref{eq13}) by fixing the sample index on the exponential term as $k_{\mathrm{c}}^{\check{m}}=(2\hat{\ell}_{0}^{\check{m}}+K_{\mathrm{pre}}-\ell_{\mathrm{D}})/2$ (the center of observed samples at the $\check{m}$-th frame) as
\begin{align}\label{eq25}
y[{\check{m}},k]\approx
\sum_{p=0}^{\hat{N}_{\mathrm{p}}-1}&\sqrt{\mathcal{E}_s}s[k-\hat{\ell}_p^{\check{m}}]h_p e^{j2\pi\nu_{p}^{\check{m}}(k_{\mathrm{c}}^{\check{m}}+{\check{m}}K)T_{\mathrm{s}}}+z[{\check{m}},k],
\end{align}
for $k=\hat{\ell}_{0}^{\check{m}},\hat{\ell}_{0}^{\check{m}}+1,\cdots,K_{\mathrm{pre}}-\ell_{\mathrm{D}}+\hat{\ell}_0^{\check{m}}$. By concatenating all the received samples, we have
\begin{equation}\label{eq26}
\by_{\check{m}}=\sqrt{\mathcal{E}_s}\bS_{\check{m}}\bh_{\check{m}}+\bz_{\check{m}},
\end{equation}
where $\by_{\check{m}}$, $\bz_{\check{m}}$, and $\bS_{\check{m}}$ are defined similarly in (\ref{eq23}), and the $p$-th element of $\bh_{\check{m}}$ is written as
\begin{equation}\label{eq27}
h_{\check{m}}[p]=h_pe^{j2\pi\nu_p^{\check{m}}(k_{\mathrm{c}}^{\check{m}}+{\check{m}}K)T_{\mathrm{s}}}.
\end{equation}
The LSE solution for linear equation in (\ref{eq26}) is
\begin{equation}\label{eq28}
\hat{\bh}_{\check{m}}=\frac{(\bS_{\check{m}}^\mathrm{H}\bS_{\check{m}})^{-1}\bS_{\check{m}}^\mathrm{H}\by_{\check{m}}}{\sqrt{\mathcal{E}_s}}.
\end{equation}

\subsection{Doppler shift estimation with phase wrapping compensation}
Using $\hat{\bh}_0$ and $\hat{\bh}_{\check{m}}$ from the previous subsections, we can obtain the Doppler shifts of the $\check{m}$-th frame as
\begin{equation}\label{eq29}
\hat{\mathrm{\nu}}_{p}^{\check{m}}=\frac{\angle(\hat{h}_{\check{m}}[p]/\hat{h}_0[p])}{2\pi (k_{\mathrm{c}}^{\check{m}}+{\check{m}}K)T_{\mathrm{s}}},
\end{equation}
for $p=0,1,\cdots,\hat{N}_{\mathrm{p}}-1$. The estimated Doppler shifts in (\ref{eq29}), however, may not be accurate for small $\check{m}$, e.g., $\nu_p^0\not\approx\hat{\nu}_p^0$, due to the approximation used in (\ref{eq21}). Instead, we exploit Doppler shift difference to obtain the Doppler shifts even for small $\check{m}$. Here, the Doppler shift difference between any consecutive frames is assumed to be fixed considering the constant vehicular velocity during one CPI. To compute the difference, we employ two properly separated frames as
\begin{equation}\label{eq30}
\boldsymbol{\mathrm{\Delta}}_{\hat{\nu}}^{\check{m},i}=\frac{\hat{\boldsymbol{\mathrm{\nu}}}^{\check{m}}-\hat{\boldsymbol{\mathrm{\nu}}}^{{\check{m}}-i}}{i},
\end{equation}
where $\hat{\boldsymbol{\nu}}^{\check{m}}\in\mathbb{R}^{\hat{N}_{\mathrm{p}}\times 1}$ is the estimated Doppler shift vector of the ${\check{m}}$-th frame with its $p$-th element defined in (\ref{eq29}).

\subsubsection{Proper value of $\check{m}$} When selecting the frame index $\check{m}$, we need to minimize the approximation error in (\ref{eq21}). Note that in (\ref{eq21}), we considered the approximation $e^{j2\pi\nu_p^0kT_{\mathrm{s}}}\approx 1$ or $2\pi\nu_p^0kT_{\mathrm{s}}\approx 0$ in the $0$-th frame while $2\pi\nu_p^0kT_{\mathrm{s}}$ would not be zero but a small value in practice. To minimize the approximation error, the phases of radar channels at the $(\check{m}-i)$-th frame must be quite larger than those in the $0$-th frame, i.e., $ \frac{ 2\pi \nu_p^{\check{m}-i}(k_{\mathrm{c}}^{\check{m}-i}+(\check{m}-i)K)T_{\mathrm{s}} }{ 2\pi \nu_p^0k_{\mathrm{c}}^{0}T_{\mathrm{s}} } \gg \epsilon$ with an arbitrary small value $\epsilon$. From this, we numerically select $\check{m}$ satisfying $\check{m}\gg i+ \frac{1}{K}\left( \frac{\nu_p^0}{\nu_p^{\check{m}-i}}k_{\mathrm{c}}^0-k_{\mathrm{c}}^{\check{m}-i} \right)$.

\subsubsection{Proper value of $i$} In (\ref{eq30}), similar to the situation of selecting proper $\check{m}$, the frame gap $i$ must be carefully determined. With large $i$, the estimation error can be reduced; however, it may induce different numbers of wrapping on the phases of $\hat{\nu}_p^{\check{m}}$ and $\hat{\nu}_p^{{\check{m}}-i}$. The phase wrapping would be significant especially when the vehicle experiences high Doppler shift due to fast velocity since the range of phase $\angle(\hat{h}_{\check{m}}[p]/\hat{h}_0[p])$ in (\ref{eq29}) is restricted to $[-\pi,\pi]$. With not-so-large $i$, the numbers of wrapping happened at the $\check{m}$-th and $(\check{m}-i)$-th frames would be the same or may be different by at most $\pm 1$. We first check whether they are different and perform preprocessing to make them the same, which is a novel technique compared to \cite{HanGLOBECOM:2020} and \cite{HanICASSP:2021} that just assume the numbers of wrapping in the two frames are the same. Then, phase wrapping compensation is conducted to unwrap the phase, which will be discussed shortly.

\begin{table*}[ht]
	\caption{Phase wrapping cases in (\ref{eq34}).}
	\centering
	\begin{tabular}{p{0.2\linewidth}| p{0.75\linewidth}}
		\hline
		Case 1 $(\psi_p<0,\,\psi_{p,\textrm{wrap}}>0)$ & $c_p=(-2\pi M_p-\psi_{p,\textrm{wrap}}^{\check{m}})D_{\check{m}}-(-2\pi M_p-\psi_{p,\textrm{wrap}}^{\check{m}-i})D_{\check{m}-i}=2\pi M_p(D_{\check{m}-i}-D_{\check{m}})-\underbrace{(\psi_{p,\textrm{wrap}}^{\check{m}}D_{\check{m}}-\psi_{p,\textrm{wrap}}^{\check{m}-i}D_{\check{m}-i})}_{=\hat{c}_p}$\\
		Case 2 $(\psi_p>0,\,\psi_{p,\textrm{wrap}}<0)$ & $c_p=(2\pi M_p+\psi_{p,\textrm{wrap}}^{\check{m}})D_{\check{m}}-(2\pi M_p+\psi_{p,\textrm{wrap}}^{\check{m}-i})D_{\check{m}-i}=-2\pi M_p(D_{\check{m}-i}-D_{\check{m}})-\underbrace{(-\psi_{p,\textrm{wrap}}^{\check{m}}D_{\check{m}}+\psi_{p,\textrm{wrap}}^{\check{m}-i}D_{\check{m}-i})}_{=\hat{c}_p}$\\
		Case 3 $(\psi_p>0,\,\psi_{p,\textrm{wrap}}>0)$ & $c_p=(2\pi M_p+\psi_{p,\textrm{wrap}}^{\check{m}})D_{\check{m}}-(2\pi M_p+\psi_{p,\textrm{wrap}}^{\check{m}-i})D_{\check{m}-i}=-2\pi M_p(D_{\check{m}-i}-D_{\check{m}})+\underbrace{(\psi_{p,\textrm{wrap}}^{\check{m}}D_{\check{m}}-\psi_{p,\textrm{wrap}}^{\check{m}-i}D_{\check{m}-i})}_{=\hat{c}_p}$ \\
		Case 4 $(\psi_p<0,\,\psi_{p,\textrm{wrap}}<0)$ & $c_p=(-2\pi M_p-\psi_{p,\textrm{wrap}}^{\check{m}})D_{\check{m}}-(-2\pi M_p-\psi_{p,\textrm{wrap}}^{\check{m}-i})D_{\check{m}-i}$\\
		& \quad\,\,$=2\pi M_p(D_{\check{m}-i}-D_{\check{m}})+\underbrace{(-\psi_{p,\textrm{wrap}}^{\check{m}}D_{\check{m}}+\psi_{p,\textrm{wrap}}^{\check{m}-i}D_{\check{m}-i})}_{=\hat{c}_p}$ \\
		\hline
	\end{tabular}
\end{table*}\label{table1}

\subsubsection{Preprocessing before phase wrapping compensation}
Assuming $\hat{\nu}_p^{{\check{m}}-i}$ experiences one more wrapping than $\hat{\nu}_p^{\check{m}}$, the Doppler shift difference becomes
\begin{align}\label{eq31}
\hat{\nu}_p^{\check{m}}-\hat{\nu}_p^{\check{m}-i}&=\psi_{p,\mathrm{wrap}}^{\check{m}}D_{\check{m}}-\psi_{p,\mathrm{wrap}+1}^{\check{m}-i}D_{\check{m}-i}\notag\\
&=\psi_{p,\mathrm{wrap}}^{\check{m}}D_{\check{m}}-(\psi_{p,\mathrm{wrap}}^{\check{m}-i}-2\pi)D_{\check{m}-i}\notag\\
&=\underbrace{(\psi_{p,\mathrm{wrap}}^{\check{m}}D_{\check{m}}-\psi_{p,\mathrm{wrap}}^{\check{m}-i}D_{\check{m}-i})}_{\mathrm{non-dominant}}+\underbrace{2\pi D_{\check{m}-i}}_{\mathrm{dominant}}.
\end{align}
In (\ref{eq31}), $\psi_{p,\mathrm{wrap}}^{\check{m}}$ denotes the residual phase after the wrapping happened in $\hat{\nu}_p^{\check{m}}$, $D_{\check{m}}$ is the inverse of denominator in (\ref{eq29}), and $\psi_{p,\mathrm{wrap}+1}^{\check{m}-i}$ denotes the residual phase after one more wrapping happened in $\hat{\nu}_p^{\check{m}-i}$. If $\hat{\nu}_p^{\check{m}}-\hat{\nu}_p^{\check{m}-i}>\sigma_{\mathrm{wrap}}$ is satisfied with a certain threshold $\sigma_{\mathrm{wrap}}$, we assume one more wrapping happened in $\hat{\nu}_p^{\check{m}-i}$ (when the dominant scatterer is moving toward to the RSU), and it is replaced as $\hat{\nu}_p^{\check{m}-i}=\psi_{p,\mathrm{wrap}}^{\check{m}-i}D_{\check{m}-i}$ after compensating this additional wrapping by
\begin{equation}\label{eq55}
\psi_{p,\mathrm{wrap}}^{\check{m}-i}=\psi_{p,\mathrm{wrap}+1}^{\check{m}-i}+2\pi.
\end{equation}
On the contrary, when the dominant scatterer is moving away from the RSU, $\hat{\nu}_p^{\check{m}}$ can have one more wrapping than $\hat{\nu}_p^{\check{m}-i}$, which can be verified if $\hat{\nu}_p^{\check{m}}-\hat{\nu}_p^{\check{m}-i}<-\sigma_{\mathrm{wrap}}$. In this case, the dominant term in (\ref{eq31}) becomes $-2\pi D_{\check{m}}$. Considering these two cases, $\sigma_{\mathrm{wrap}}$ should be smaller than $\pi D_{\check{m}}$ and $\pi D_{\check{m}-i}$ due to possible opposite signs of non-dominant and dominant terms. Thus, we set the threshold $\sigma_{\mathrm{wrap}}=\pi (D_{\check{m}}+D_{\check{m}-i})/2$. We will show in Fig. \ref{nu_diff_NMSE} in Section VI that this preprocessing works well.

\subsubsection{Phase wrapping compensation}
Assuming $i$ is not-so-large such that $\hat{\nu}_p^{\check{m}}$ and $\hat{\nu}_p^{\check{m}-i}$ experience the same number of phase wrapping through preprocessing, the RSU needs to figure out how many times the wrapping occurred in $\angle(\hat{h}_{\check{m}}[p]/\hat{h}_0[p])$ of (\ref{eq29}). To do this, we adopt the wrapping corrector defined as
\begin{equation}\label{eq32}
\hat{c}_p=|\hat{\nu}_{p}^{\check{m}}|-|\hat{\nu}_{p}^{{\check{m}}-i}|,
\end{equation}
for $p=0,1,\cdots,\hat{N}_{\mathrm{p}}-1$. The wrapping corrector for the true phases is
\begin{align}\label{eq33}
c_p&=|\nu_p^{\check{m}}|-|\nu_p^{{\check{m}}-i}|\notag\\
&=|\psi_p^{\check{m}} D_{\check{m}}|-|\psi_p^{{\check{m}}-i} D_{{\check{m}}-i}|\notag\\
&=|(2\pi M_p+\psi_{p,\mathrm{wrap}}^{{\check{m}}})D_{\check{m}}|-|(2\pi M_p+\psi_{p,\mathrm{wrap}}^{{\check{m}}-i})D_{{\check{m}}-i}|,
\end{align}
where $\psi_p^{\check{m}}$ represents the true phase (in radian) for the $p$-th dominant scatterer of the ${\check{m}}$-th frame, and $M_p$ is the number of wrapping happened on the phase of Doppler shifts at the $\check{m}$-th and $(\check{m}-i)$-th frames. It is clear from (\ref{eq33}) that
\begin{align}\label{eq34}
M_p=\begin{cases}\frac{\hat{c}_p}{2\pi (D_{{\check{m}}-i}-D_{\check{m}})}+\frac{c_p}{2\pi (D_{{\check{m}}-i}-D_{\check{m}})},&\mathrm{for}\quad\mathrm{Case\,1}\\
-\frac{\hat{c}_p}{2\pi (D_{{\check{m}}-i}-D_{\check{m}})}-\frac{c_p}{2\pi (D_{{\check{m}}-i}-D_{\check{m}})},&\mathrm{for}\quad\mathrm{Case\,2}\\
\frac{\hat{c}_p}{2\pi (D_{{\check{m}}-i}-D_{\check{m}})}-\frac{c_p}{2\pi (D_{{\check{m}}-i}-D_{\check{m}})},&\mathrm{for}\quad\mathrm{Case\,3}\\
-\frac{\hat{c}_p}{2\pi (D_{{\check{m}}-i}-D_{\check{m}})}+\frac{c_p}{2\pi (D_{{\check{m}}-i}-D_{\check{m}})},&\mathrm{for}\quad\mathrm{Case\,4}
\end{cases},
\end{align}
where the four cases are derived in TABLE II. Although it is not possible to know $c_p$ in practice, this term is to make $M_p$ an integer value. Therefore, we adopt approximated $M_p$ as
\begin{align}\label{eq35}
\bar{M}_p\approx\begin{cases}\lfloor\frac{\hat{c}_p}{2\pi (D_{{\check{m}}-i}-D_{\check{m}})}\rceil,&\mathrm{for}\quad\psi_{p,\mathrm{wrap}}>0\\
\lfloor-\frac{\hat{c}_p}{2\pi (D_{{\check{m}}-i}-D_{\check{m}})}\rceil,&\mathrm{for}\quad\psi_{p,\mathrm{wrap}}<0
\end{cases}.
\end{align}
Using $\bar{M}_p$, the phase wrapping is compensated as
\begin{equation}\label{eq36}
\hat{\hat{\nu}}_{p}^{\check{m}}=\hat{\nu}_{p}^{\check{m}}+2\pi\bar{M}_p D_{\check{m}},\quad \hat{\hat{\nu}}_{p}^{{\check{m}}-i}=\hat{\nu}_{p}^{{\check{m}}-i}+2\pi\bar{M}_p D_{{\check{m}}-i}.
\end{equation}
The Doppler shift difference $\boldsymbol{\Delta}_{\hat{\hat{\nu}}}^{\check{m},i}$ in (\ref{eq30}) is correctly obtained using the refined estimates $\hat{\hat{\nu}}_p^{\check{m}}$ and $\hat{\hat{\nu}}_p^{{\check{m}}-i}$, and the Doppler shifts of the $m$-th frame can be finally obtained as
\begin{align}\label{eq37}
\hat{\hat{\boldsymbol{\nu}}}^m=\hat{\hat{\boldsymbol{\nu}}}^{\check{m}}+(m-{\check{m}})\Delta_{\hat{\hat{\nu}},\mathrm{med}}^{\check{m},i}\mathcal{\mathbf{1}},
\end{align}
for $m=0,1,\cdots,M-1$, where $\Delta_{\hat{\hat{\nu}},\mathrm{med}}^{\check{m},i}$ is the median of all the elements of $\boldsymbol{\Delta}_{\hat{\hat{\nu}}}^{\check{m},i}$. The reason why we adopt the median will become clear in Fig. \ref{nu_diff_NMSE} of Section VI.

\section{ISAR Image Formation}
The general ISAR imaging technique is to perform the FFT of the radar echo signals passed through a matched filter along the cross-range direction \cite{Ozdemir:2012}. This conventional method can be applied to our case as well since the proposed delay estimation method can be considered as the matched filtering for the partial echoes. The conventional method, however, provides ISAR images with inaccurate cross-range scaling as shown in Section VI since the IEEE 802.11ad waveform is not optimized for pulse compression, which requires the frequency variation of transmit signals in time. Thus, we propose a novel ISAR imaging technique to obtain accurate ISAR images by forming an image with range and cross-range profiles.

A range profile can be simply obtained from the correlation function defined in (\ref{eq17}) by transforming $\ell$ into the range bin. Then, the range and cross-range profiles are intertwined with each other to form a pre-image matrix $\bP_{\mathrm{I}}$, which will be processed to formulate an ISAR image. The RSU finally obtains ISAR images with the same cross-range scaling through performing the FFT of the pre-image along the cross-range direction.

Before constructing the cross-range profile, we first require the information of rotational velocity $\omega$. This can be obtained with the knowledge of vehicular velocity $V_{\mathrm{X}}$, where $\omega=V_{\mathrm{X}}^{\perp}/R_0$ with the cross-range component of velocity $V_{\mathrm{X}}^{\perp}=V_{\mathrm{X}}\mathrm{cos}(\mathrm{tan}^{-1}(X_0/Y_0))$ and the distance between vehicle and RSU $R_0=\sqrt{X_0^2+Y_0^2+Z_0^2}$. The vehicular velocity $V_{\mathrm{X}}$ is obtained from the relationship between Doppler shift estimates and equation of motion, which are given as
\begin{align}\label{eq38,39,40}
\nu_{p}^0&=\frac{2V_{\mathrm{X}} \mathrm{sin}(\phi_p^0)}{\lambda}\approx\frac{2V_{\mathrm{X}} \phi_p^0}{\lambda},\\
\nu_{p}^{M-1}&=\frac{2V_{\mathrm{X}} \mathrm{sin}(\phi_p^{M-1})}{\lambda}\approx\frac{2V_{\mathrm{X}} \phi_p^{M-1}}{\lambda},\\
\mathrm{CPI}\cdot V_{\mathrm{X}}&=R_0 (\mathrm{sin}(\phi_p^0)-\mathrm{cos}(\phi_p^0)\mathrm{tan}(\phi_p^{M-1}))\notag\\
&\approx R_0(\phi_p^0-\phi_p^{M-1}),
\end{align}
where all the approximations are from the small-angle approximation due to the relatively long distance between the RSU and vehicle. The above relation would hold for any $p$ since all the dominant scatterers on the vehicle move with the same velocity $V_{\mathrm{X}}$. From (40)-(42), the vehicular velocity is estimated as
\begin{equation}\label{eq41}
\hat{V}_{\mathrm{X}}\approx\sqrt{\frac{\lambda R_0(\hat{\hat{\nu}}_{p}^0-\hat{\hat{\nu}}_{p}^{M-1})}{2\cdot\mathrm{CPI}}}.
\end{equation}

\begin{figure}
	\centering
	\includegraphics[width=0.99\columnwidth]{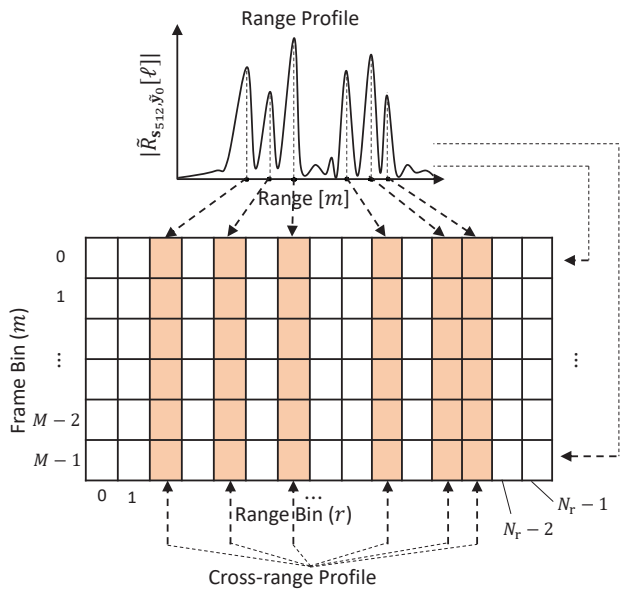}
	\caption{An example of pre-image matrix formation procedure.}\label{image_form}
\end{figure}

Using $\omega$, the Doppler shift estimates $\hat{\hat{\nu}}_p^m$ are directly transformed into the cross-range information $\hat{\hat{\nu}}_p^mc/(2f_{\mathrm{c}}\omega\Delta_{\mathrm{cr}})$, where the cross-range resolution is $\Delta_{\mathrm{cr}}=\lambda W_{\mathrm{D}}/(2M\omega)$ with the Doppler frequency bandwidth $W_{\mathrm{D}}=2\omega Y_{\mathrm{size}}f_{\mathrm{c}}/c$ and the image projection plane size $X_{\mathrm{size}}\times Y_{\mathrm{size}}$, and the velocity resolution $\Delta_{\mathrm{v}}=\lambda\Delta_{\mathrm{D}}/2$ is obtained from the Doppler resolution $\Delta_{\mathrm{D}}=W_{\mathrm{D}}/M$ \cite{Ozdemir:2012}. From the obtained cross-range information, we define the cross-range profile $\mathrm{CR}_p[m]=e^{j\frac{2\pi}{N_{\mathrm{cr}}}m\hat{\hat{\nu}}_p^m\frac{c}{2f_{\mathrm{c}}\omega\Delta_{\mathrm{cr}}}}$ for the $p$-th dominant scatterer, which determines the $m$-th cross-range position, where $N_{\mathrm{cr}}=\lfloor{Y_{\mathrm{size}}/\Delta_{\mathrm{cr}}\rfloor}=M$ is the number of cross-range bins. Then, with the number of range bins $N_{\mathrm{r}}=\lfloor{X_{\mathrm{size}}/\Delta_{\mathrm{r}}\rfloor}$, the $(m,r)$-th element of pre-image matrix $\bP_{\mathrm{I}}\in\mathbb{C}^{N_{\mathrm{cr}}\times N_{\mathrm{r}}}$ for one CPI with $M$ frames is obtained as
\begin{equation}\label{eq42}
P_{\mathrm{I}}[m,r]=
\begin{cases}
\left|\tilde{R}_{\bs_{512}\tilde{\by}_0}[\hat{\ell}_p^0]\right|\mathrm{CR}_p[m],&\mathrm{if}\,\, (\mathrm{C}^*)\\
\left|\tilde{R}_{\bs_{512}\tilde{\by}_0}\left[\lceil\frac{2}{cT_{\mathrm{s}}}(r\Delta_{\mathrm{r}}+R_0-\frac{X_{\mathrm{size}}}{2})\rceil\right]\right|,&\mathrm{else}
\end{cases},
\end{equation}
with the condition in (\ref{eq42}) given as
\begin{equation}
(\mathrm{C}^*): -\Delta_{\mathrm{r}}<r\Delta_{\mathrm{r}}-\frac{c\hat{\ell}_p^0 T_{\mathrm{s}}}{2}+R_0-\frac{X_{\mathrm{size}}}{2}<0,
\end{equation}
where $\Delta_{\mathrm{r}}=c/(2W)$ is the range resolution with the bandwidth $W$. The condition $(\mathrm{C}^*)$ represents the case that the distance to the $p$-th dominant scatterer is in the $r$-th range bin. The pre-image matrix $\bP_{\mathrm{I}}$ is a preliminary image having the corresponding cross-range information in the colored range bins where the dominant scatterers exist as shown in Fig. \ref{image_form}. Finally, the ISAR image is formed through the FFT of the pre-image matrix along the cross-range direction.

Remark 1: In our scenario, the range migration, which usually happens for the ISAR imaging of targets with extremely high velocity like airplanes \cite{Ozdemir:2012}, can be ignored due to the use of short CPI and relatively slower speed of vehicles. Also, for the fine range alignment, the proposed ISAR imaging technique only employs the range profile of $0$-th frame in Section VI.

Remark 2: To compensate the effect of time-varying Doppler shifts due to the translational motion, which results in blurred ISAR images, the technique called phase adjustment is required in general \cite{Wang:2004}. The proposed ISAR imaging technique, however, does not require this process since it exploits the estimated Doppler shifts of all frames, as detailed in Section IV, not the echo signals directly to form ISAR images.

Remark 3: As a further discussion, we briefly look into multi-vehicle environments. For the case of multiple vehicles, there are two possible scenarios: i) some large vehicles like trucks may act as dominant clutters, or ii) the RSU needs to form the image of multiple vehicles simultaneously. Even though the typical ISAR imaging is for a single target, the delays and Doppler shifts still can be estimated using the multi-target model for multiple vehicles. Since different vehicles may have different velocities, it would be quite easy to handle the first scenario, i.e., get rid of some dominant scatterers that have inconsistent Doppler shifts compared to those of the target vehicle. Note that for the typical ISAR imaging scenario including the one in Fig. \ref{V2I}, all dominant scatterers are on a single target vehicle, causing the same velocity for all dominant scatterers. The second scenario that needs to perform the ISAR imaging with the inconsistent Doppler shifts, however, is not an easy task since different velocities make the phase wrapping compensation quite difficult. Therefore, more advanced techniques are required to develop the ISAR imaging for multiple vehicles, which is an interesting future research problem.

\begin{figure}
	\centering
	\includegraphics[width=0.653\columnwidth]{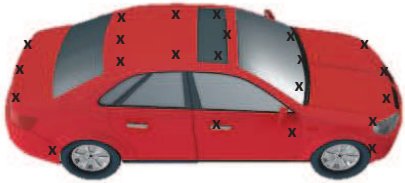}
	\caption{A point scatterer model based on the shape of a realistic vehicle for $X_0=$ 0 m (Side view).}\label{scat_model_1}
\end{figure}

\begin{figure}
	\centering
	\includegraphics[width=0.9\columnwidth]{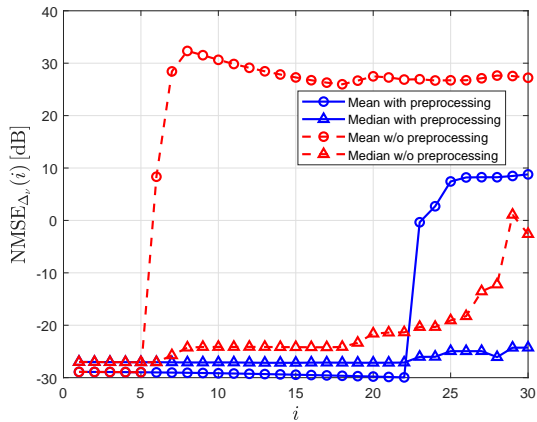}
	\caption{NMSEs of Doppler shift difference in accordance with frame gap $i$ in (\ref{eq30}) when $\check{m}=M-1$ and CPI = 2 ms.}\label{nu_diff_NMSE}
\end{figure}

\section{Simulation Results}
We perform simulations with realistic vehicular models to evaluate the proposed ISAR imaging based on the IEEE 802.11ad waveform. The TX (for both radar and communications) and RX (for radar) antennas at the RSU are $8\times8$ UPAs, and the RX antennas at the vehicle are an $8\times2$ UPA. The Rician $K$-factor $K_{\mathrm{c}}$ is assumed to be 12.347 dB \cite{Mukherjee:2017}. Note that the RX antennas at the vehicle and $K_{\mathrm{c}}$ are used to find the beamformers in (\ref{eq10}). From the IEEE 802.11ad specification, we adopt the bandwidth $W=$ 1.76 GHz, carrier frequency $f_{\mathrm{c}}=$ 60 GHz, root-raised cosine (RRC) filter using roll-off factor 0.25 for the pulse shaping filter and matched filter, training samples in a frame $K_{\mathrm{pre}}=3328$, and time period of one frame $T_{\mathrm{f}}=KT_{\mathrm{s}}$ with $T_{\mathrm{s}}=1/W$ and\footnote{In the IEEE 802.11ad SC PHY, the lengths of STF, CEF, and a header are fixed while the length of BLKs is varied according to the required amount of data for the target vehicle. We choose the minimum number of samples $K$ in a frame to show that the proposed ISAR imaging is possible even for short CPI. This is essential for the proposed ISAR imaging technique to be useful for high speed vehicular environments.} $K=13632$ \cite{Spec:2012}. The RSU is located at the origin while the initial location of vehicle is $(Y_0, Z_0) =$ (20 m, -7 m) and $X_0$ varies to evaluate the proposed ISAR imaging with different points of view. The velocity and size of vehicle are $V_{\mathrm{X}}=$ 40 m/s and $(X_{\mathrm{v}}, Y_{\mathrm{v}}, Z_{\mathrm{v}}) =$ (4.8 m, 2.1 m, 1.5 m). We also assume the path-loss exponent 2, RCS of a vehicle $\sigma_{\mathrm{RCS}}=$ 20 dBsm, RCS per dominant scatterer in a linear scale $\sigma_p^{\mathrm{RCS}}=10^{(\sigma_{\mathrm{RCS}}/10)}/N_{\mathrm{p}}$ m$^2$, the number of frames $M=\lfloor\mathrm{CPI}/T_{\mathrm{f}}\rfloor$. We set the design parameter $\ell_{\mathrm{D}}=1$ to avoid the interference from the echoes of data streams as mentioned in Section II-C and to use the received signals maximally in (\ref{eq22}) and (\ref{eq26}) when obtaining the backscattering coefficients and effective radar channel estimates. The number of dominant scatterers $N_{\mathrm{p}}$ varies in each point scatterer model depending on $X_0$. The signal power transmitted at the RSU is 30 dBm. The power incorporating the noise and clutter is assumed to be $\sigma_{\mathrm{nc}}^2=N_0W+P_{\mathrm{c}}$ with the noise spectral density $N_0=-174$ dBm/Hz and the average clutter power $P_{\mathrm{c}}$, which is obtained as
\begin{equation}
P_{\mathrm{c}}=\frac{ \mathcal{E}_s G_a^2\lambda^2\sigma_s }{ (4\pi)^3R_0^4\mathrm{cos}\left( \mathrm{tan}^{-1}\left( \frac{Z_0}{Y_0} \right) \right) },
\end{equation}
where $G_a=\pi/(\theta_B\phi_B)$ is the antenna gain with the azimuth and elevation beam widths $\phi_B$ and $\theta_B$, and $\sigma_s=\frac{\sigma_0\theta_B R_0cT_{\mathrm{s}}}{2}$ is the RCS per unit area with $\sigma_0=0.15\mathrm{sin}\left( \mathrm{tan}^{-1}\left(\frac{Z_0}{Y_0}\right) \right)$ \cite{lacomme:2001,skolnik:1980}. The size of image projection plane is $X_{\mathrm{size}}\times Y_{\mathrm{size}}=$ 15 m $\times$ 25 m to visualize all the images from several points of view with the same image size. We design a point scatterer model to consider a realistic vehicle at $X_0=$ 0 m as shown in Fig. \ref{scat_model_1}.

To choose proper $i$ in (\ref{eq30}) and evaluate the effectiveness of proposed preprocessing, we first show the normalized mean square error (NMSE) of Doppler shift difference according to the frame gap $i$ in Fig. \ref{nu_diff_NMSE}, which is defined as
\begin{equation}\label{eq43}
\mathrm{NMSE}_{\Delta_{\nu}}(i)=\mathbb{E}\begin{bmatrix}\begin{pmatrix}
\frac{\Delta_{\hat{\hat{\nu}},\mathrm{STAT}}^{\check{m},i}-\Delta_{\nu,\mathrm{true}}}{\Delta_{\nu,\mathrm{true}}}\end{pmatrix}^2\end{bmatrix},
\end{equation}
where $\Delta_{\nu,\mathrm{true}}$ is the true Doppler shift difference and $\mathrm{STAT}\in\left\{\mathrm{med,mean}\right\}$. From $i=6$, some dominant scatterers start to experience different numbers of phase wrapping at the $\check{m}$-th and $(\check{m}-i)$-th frames, which results in sharp increase of NMSE for the case of ``Mean w/o preprocessing.'' On the contrary, the proposed preprocessing of phase wrapping works well up to sufficiently large $i$. The figure clearly shows that using the median of Doppler shift differences $\Delta_{\hat{\hat{\nu}},\mathrm{med}}^{\check{m},i}$, as in (\ref{eq37}), is much more robust than using the mean of differences for large $i$ since the mean is heavily affected by ineffective phase wrapping compensation when some dominant scatterers experience different numbers of phase wrapping at the $\check{m}$-th and $(\check{m}-i)$-th frames. From now on, we only consider the results with preprocessing and $i=6$ for the phase wrapping compensation.

\begin{figure}
	\centering
	\includegraphics[width=0.9\columnwidth]{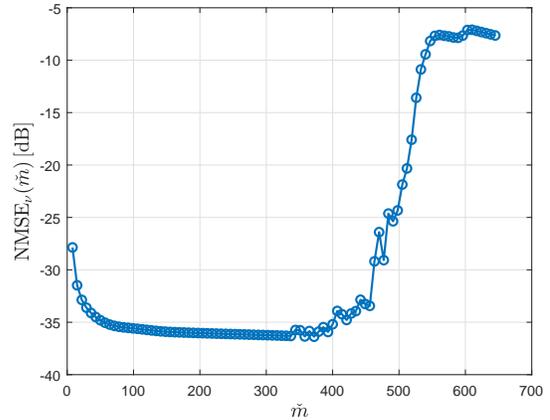}
	\caption{NMSE of Doppler shifts at the $\check{m}$-th frame when $i=6$.}\label{nu_NMSE}
\end{figure}

We show the proposed Doppler estimation technique can be extremely accurate by adopting appropriate $\check{m}$ for any CPI in Fig. \ref{nu_NMSE}. The NMSE of Doppler shift at the $\check{m}$-th frame in Fig. \ref{nu_NMSE} is defined as
\begin{equation}\label{eq44}
\mathrm{NMSE}_{\nu}(\check{m})=\frac{1}{\hat{N}_{\mathrm{p}}}\sum_{p=0}^{\hat{N}_{\mathrm{p}}-1}\mathbb{E}\begin{bmatrix}\begin{pmatrix}
\frac{\hat{\hat{\nu}}_p^{\check{m}}-\nu_p^{\check{m}}}{\nu_p^{\check{m}}}\end{pmatrix}^2\end{bmatrix}.
\end{equation}
When $\check{m}$ is too small, the NMSE is large due to the approximation error in (\ref{eq21}). When $\check{m}$ becomes large, the NMSE explodes due to inefficient phase wrapping compensation by the round-off error in the second term for each case in (\ref{eq34}). With proper values of $\check{m}$, where the range of proper value is large, the proposed Doppler shift estimation technique can achieve very small NMSE. Since $\check{m}$ is a system parameter the RSU can determine, Fig. \ref{nu_NMSE} also shows that the Doppler shift can be accurately estimated for any CPI values as long as CPI is not extremely small, e.g., the possible range of $\check{m}$ is 0 to 257 for CPI = 2 ms. In other words, the RSU can estimate high Doppler shift with very small NMSE even for short CPI. For the remaining results, we adopt CPI = 2 ms with\footnote{The RSU can transmit many frames in a row by allocating a contention-free dedicated resource to the target vehicle \cite{Mohebi:2020}.} $M=258$ while much smaller $M$, e.g., $M=50$, may be enough in practice.

\begin{figure}
	\centering
	\includegraphics[width=0.9\columnwidth]{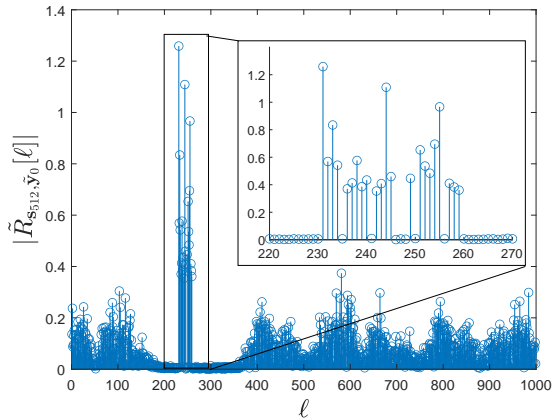}
	\caption{Delay estimation using the correlation function defined in (\ref{eq17}) for $X_0=$ 0 m.}\label{delay_est}
\end{figure}

\begin{figure}
	\centering
	\includegraphics[width=0.9\columnwidth]{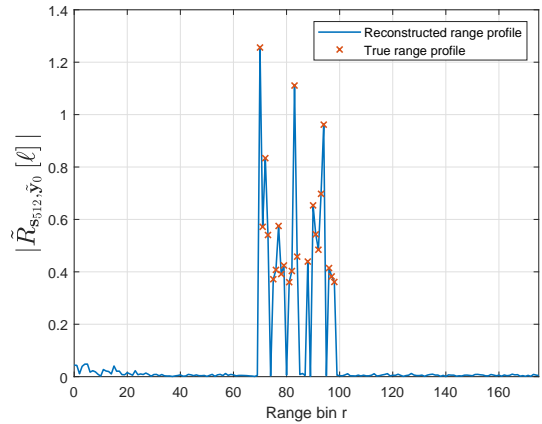}
	\caption{A range profile reconstruction through the delay estimation for $X_0=$ 0 m where $\ell=\lceil\frac{2}{cT_{\mathrm{s}}}(r\Delta_{\mathrm{r}}+R_0-\frac{X_{\mathrm{size}}}{2})\rceil$, which transforms the index $\ell$ into the range bin.}\label{range_prof}
\end{figure}

Fig. \ref{delay_est} verifies that the delays of dominant scatterers can be estimated perfectly due to the ideal auto-correlation property of Golay complementary sequences. Then, we obtain the range profile of dominant scatterers by transforming the index $\ell$ into the range bin as shown in Fig. \ref{range_prof}. The range resolution using the IEEE 802.11ad waveform is about 8.52 cm, which is enough to differentiate most of dominant scatterers distributed along the range direction.

\begin{figure*}
	\centering
	\begin{subfigure}{0.333\linewidth}
		\centering
		\includegraphics[width=\linewidth]{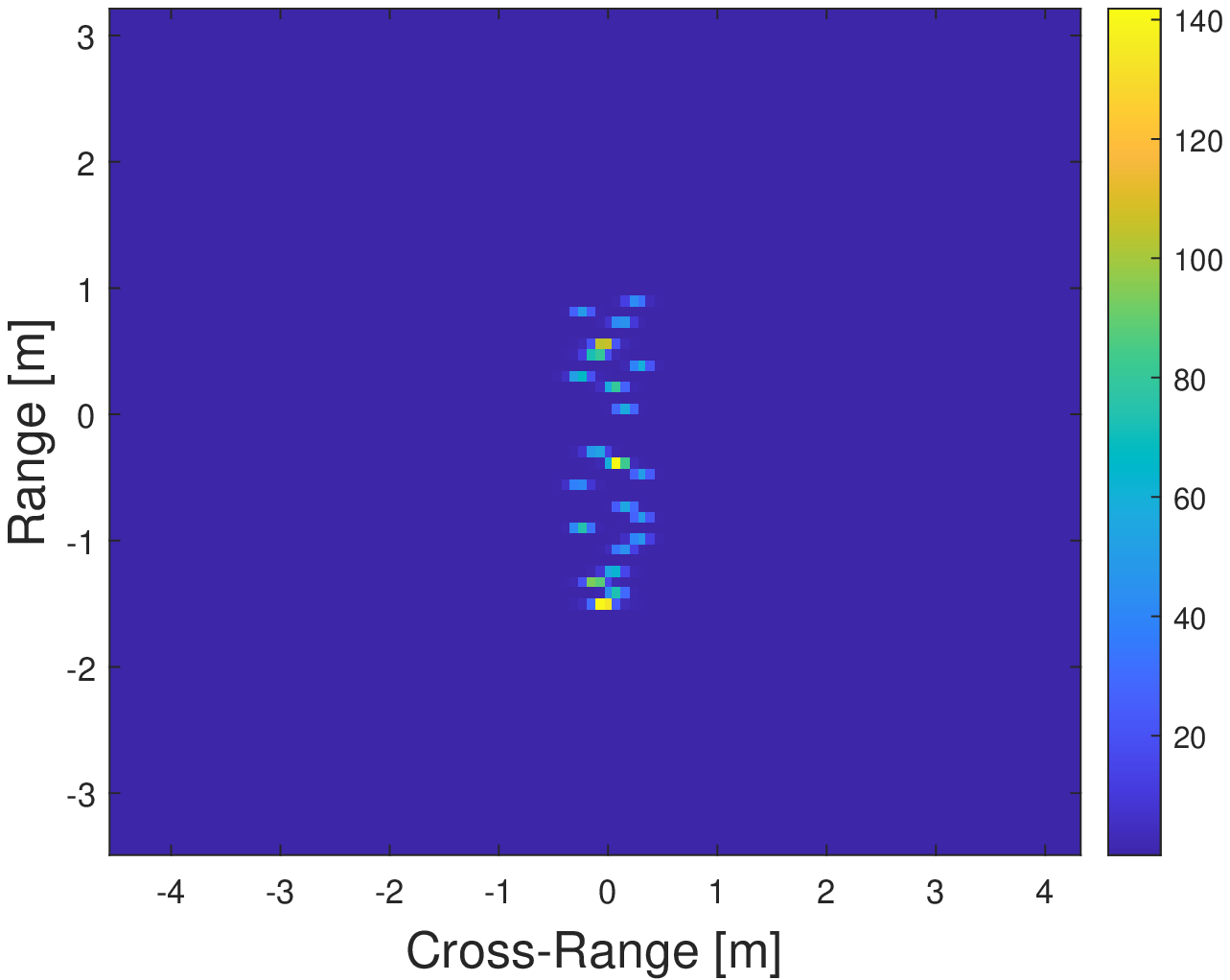}
		\caption{}
		\label{ISAR_conventional}
	\end{subfigure}\hfill
	\begin{subfigure}{0.333\linewidth}
		\centering
		\includegraphics[width=\linewidth]{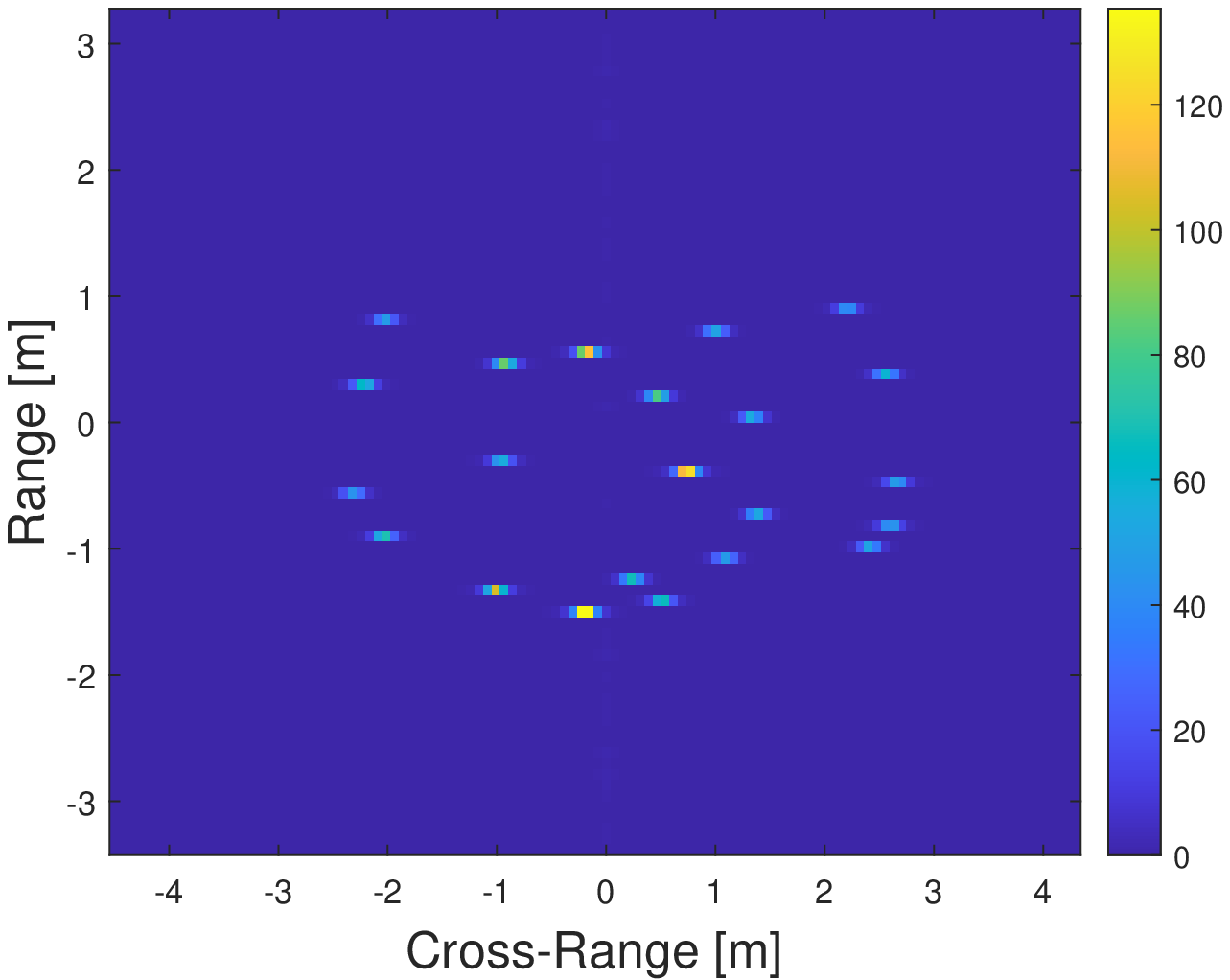}
		\caption{}
		\label{ISAR_1_flip}
	\end{subfigure}\hfill
	\begin{subfigure}{0.333\linewidth}
		\centering
		\includegraphics[width=\linewidth]{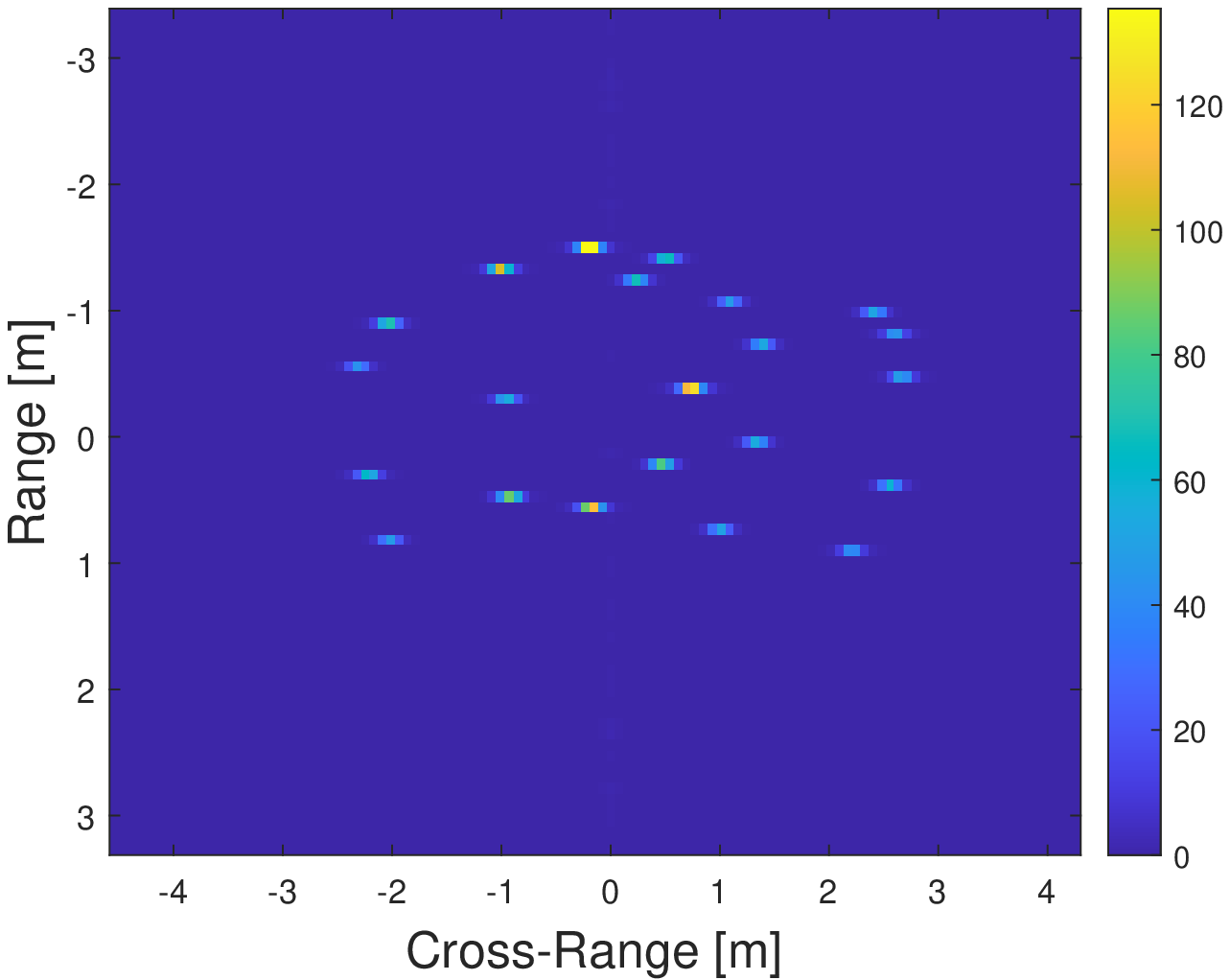}
		\caption{}
		\label{ISAR_1}
	\end{subfigure}
	\caption{The ISAR images for $X_0=$ 0 m. (a) conventional method; (b) proposed technique; (c) flipped version of (b).}
\end{figure*}

\begin{figure}
	\centering
	\begin{subfigure}{0.5\linewidth}
		\centering
		\includegraphics[width=\linewidth]{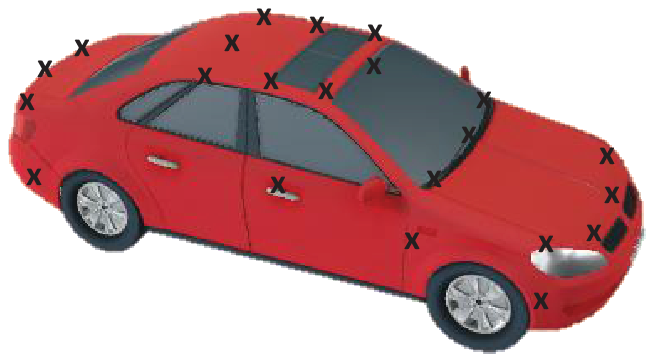}
	\end{subfigure}\hfill
	\begin{subfigure}{0.5\linewidth}
		\centering
		\includegraphics[width=\linewidth]{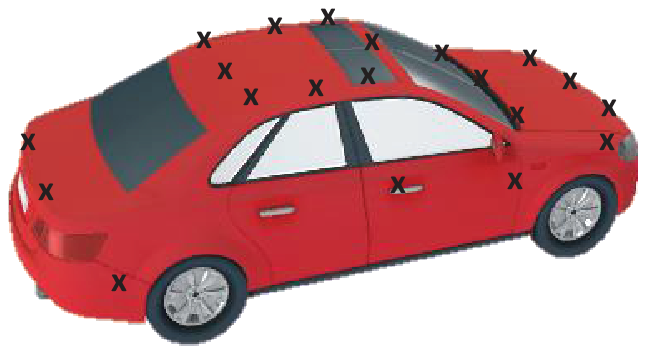}
	\end{subfigure}
	
	\vspace{3ex}
	
	\begin{subfigure}{0.5\linewidth}
		\centering
		\includegraphics[width=\linewidth]{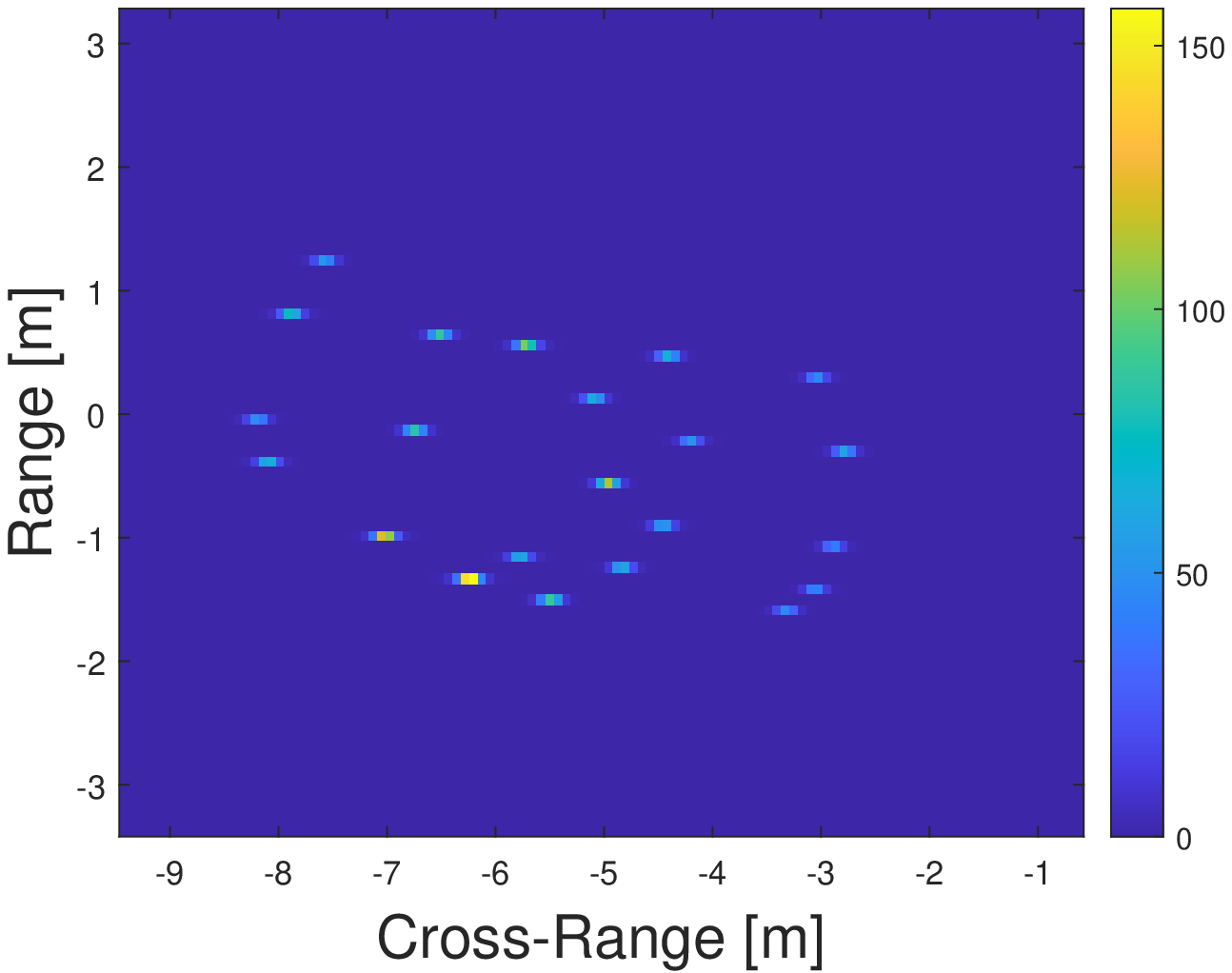}
	\end{subfigure}\hfill
	\begin{subfigure}{0.5\linewidth}
		\centering
		\includegraphics[width=\linewidth]{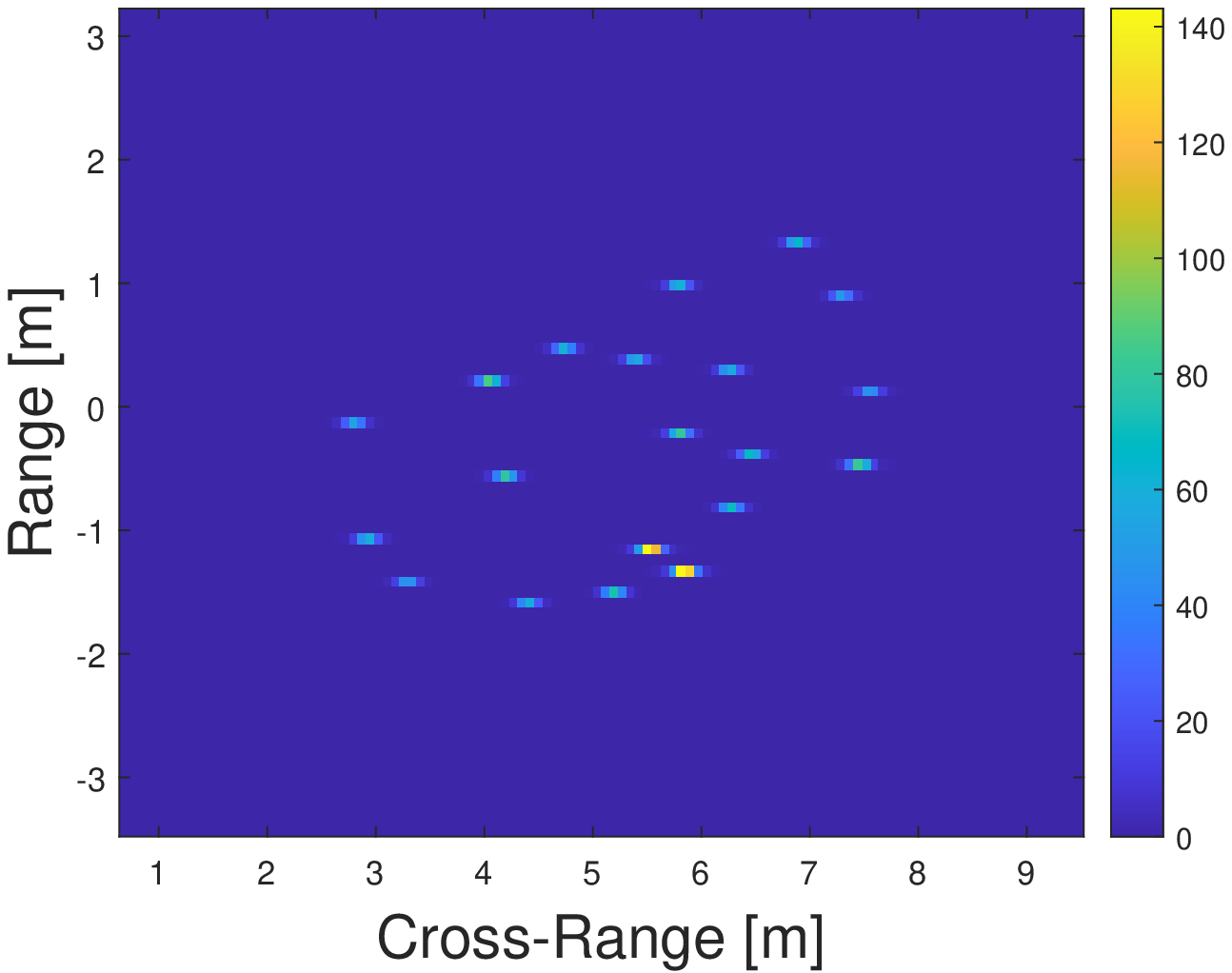}
	\end{subfigure}
	
	\vspace{3ex}
	
	\begin{subfigure}{0.5\linewidth}
		\centering
		\includegraphics[width=\linewidth]{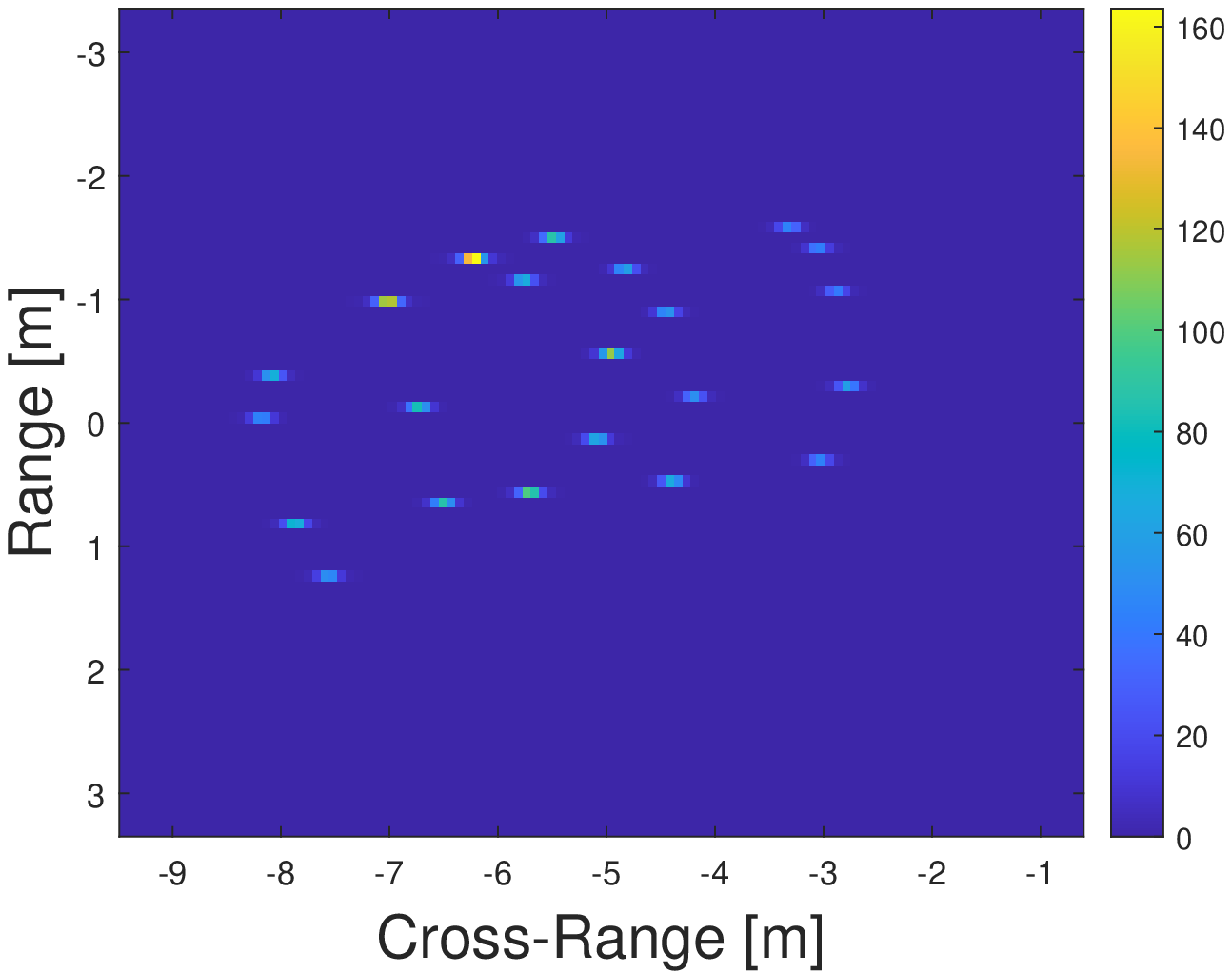}
		\caption{}
		\label{ISAR_2}
	\end{subfigure}\hfill
	\begin{subfigure}{0.5\linewidth}
		\centering
		\includegraphics[width=\linewidth]{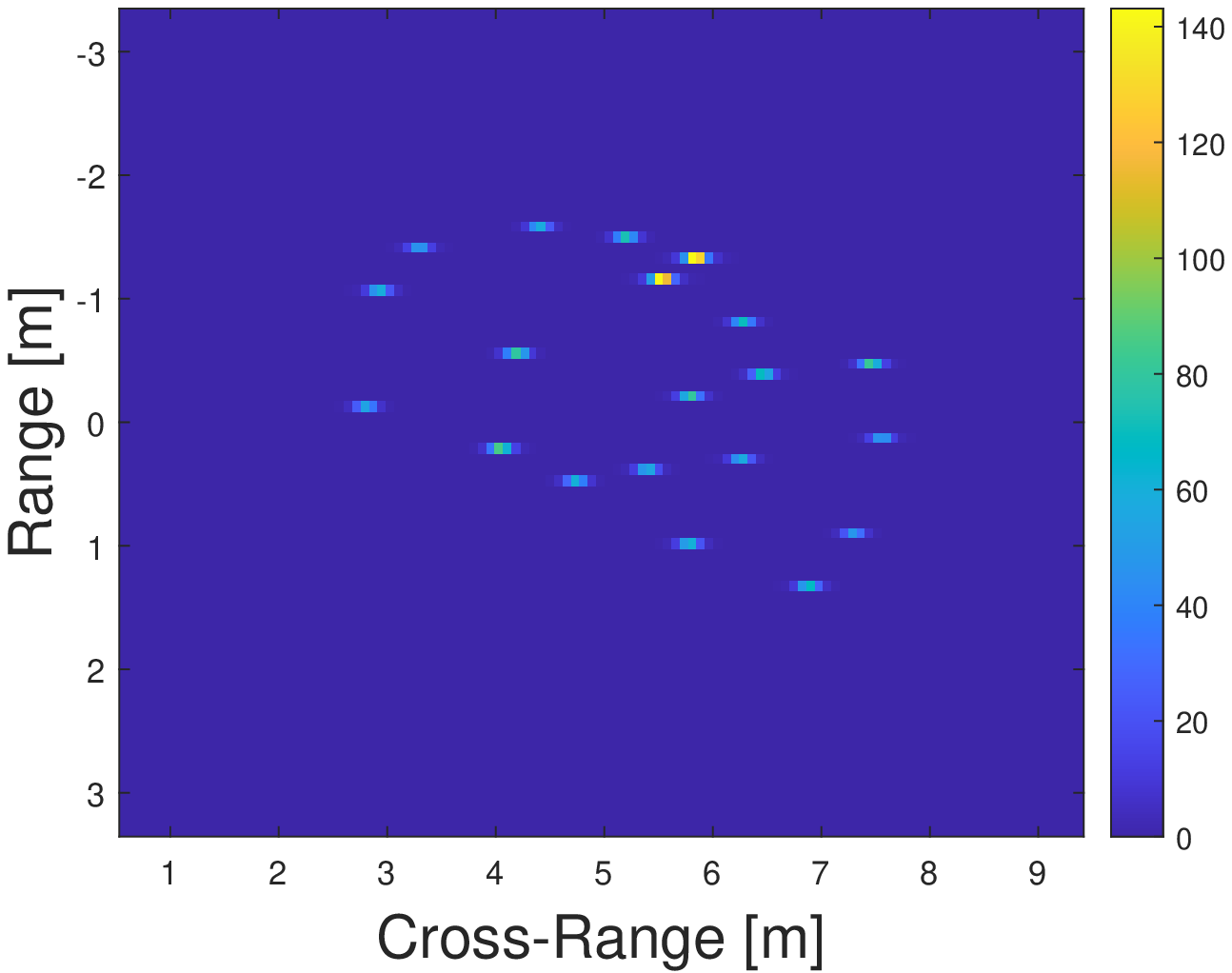}
		\caption{}
		\label{ISAR_3}
	\end{subfigure}
	
	\caption{The point scatterer models, ISAR images, and flipped versions from different points of view. (a) for $X_0=$ -5 m; (b) for $X_0=$ 5 m.}\label{ISAR_2_3}
\end{figure}

Fig. \ref{ISAR_conventional} shows the ISAR image for $X_0=$ 0 m obtained from the conventional method \cite{Ozdemir:2012}, which is to perform the FFT along the cross-range direction after the matched filtering to the partial echo signals, has inaccurate cross-range scaling. On the contrary, the ISAR image using the proposed technique in Fig. \ref{ISAR_1_flip} is well-scaled while the image is flipped since some dominant scatterers on the roof of vehicle are closer than other parts to the RSU. The typical ISAR imaging considers targets over radar systems, e.g., aircraft, while the vehicle is below the radar module at the RSU in our scenario. Therefore, the ISAR image is flipped according to the general definition of range and cross-range axes of ISAR image. By flipping the original ISAR image, we obtain Fig. \ref{ISAR_1} that exhibits well the vehicular shape as the point scatterer model in Fig. \ref{scat_model_1}. Compared to \cite{HanGLOBECOM:2020}, there is no blurring in the cross-range domain due to the novel definition of pre-image matrix in (\ref{eq42}).

\begin{figure}
	\begin{subfigure}{0.5\linewidth}
		\centering
		\includegraphics[width=\linewidth]{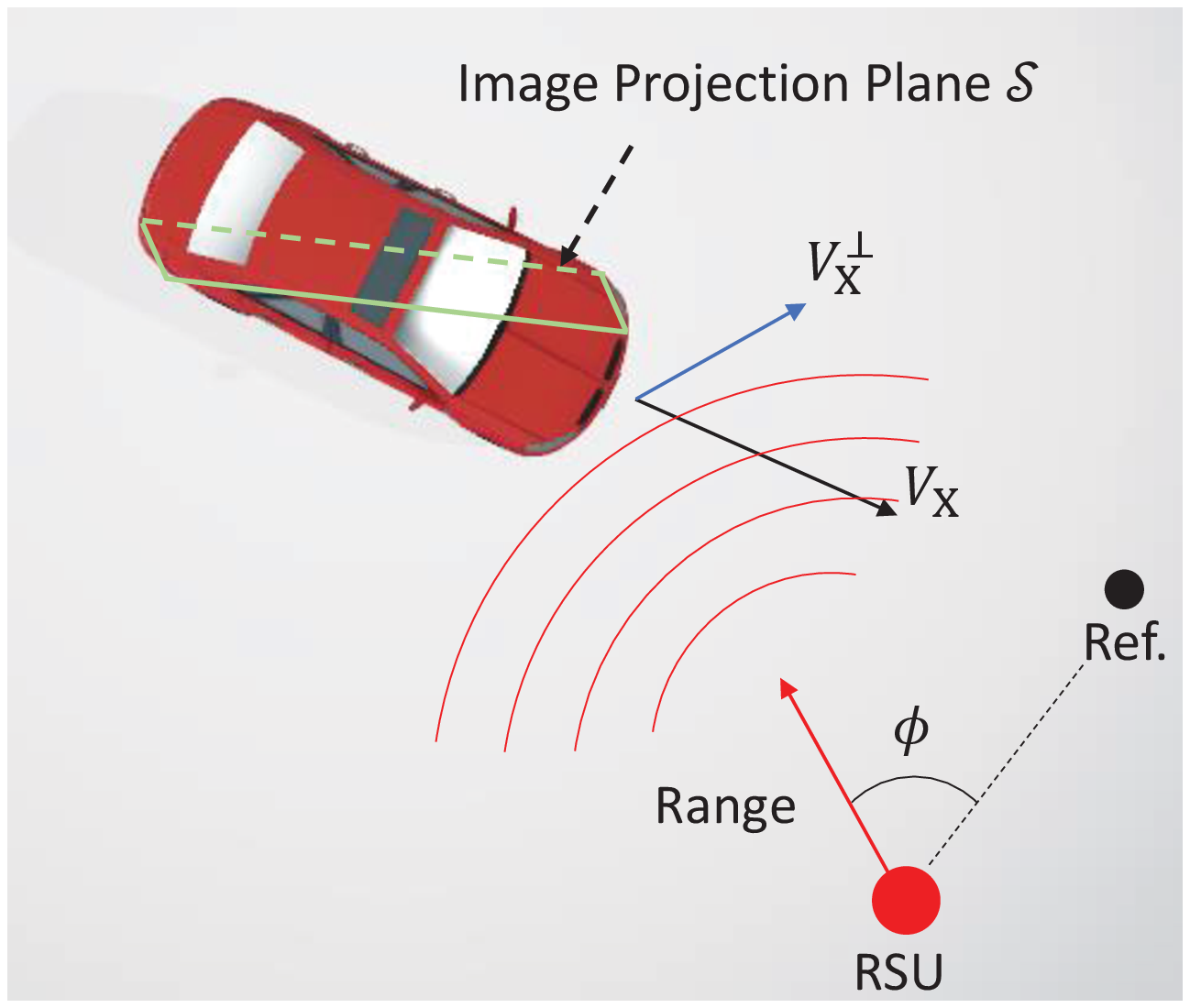}
		\caption{}
		\label{proj_1}
	\end{subfigure}\hfill
	\begin{subfigure}{0.4\linewidth}
		\centering
		\includegraphics[width=\linewidth]{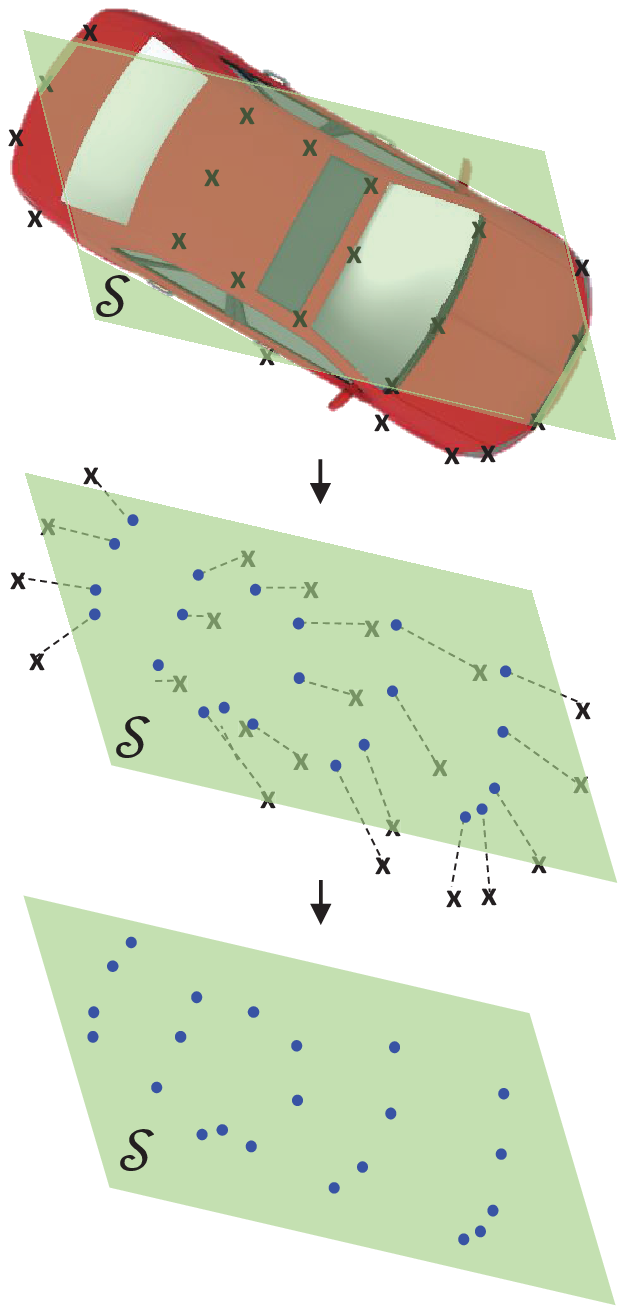}
		\caption{}
		\label{proj_2}
	\end{subfigure}
	
	\caption{An example of image projection plane for $X_0=$ -5 m. (a) image projection plane (aerial view); (b) the projection process on plane of dominant scatterers.}
\end{figure}

We show in Fig. \ref{ISAR_2_3} the point scatterer models, ISAR images of vehicle, and flipped versions of them from different points of view. While the flipped images clearly show the shape of target vehicles, the angles of direction are different from the point scatterer models due to flipping. Note that the flipped images are just to show the shape of vehicle visually while the RSU can directly process the original ISAR images, which have the same angles of direction with the point scatterer models, for advanced operations such as beam tracking because the RSU \textit{knows in advance} that the target vehicles on original ISAR images are flipped.

In Fig. \ref{ISAR_2_3}, the range on both ends of the vehicle appears to be distorted compared to the point scatterer models. The distortion from different viewpoints can be analyzed by the image projection plane. The ISAR image projection plane is determined as the two-dimensional (2D) surface including the perpendicular vehicular velocity and the beam transmission direction from the RSU to the moving vehicle as $\mathcal{S}$ in Fig. \ref{proj_1}. The distortion on both ends of the vehicle happens since the dominant scatterers on both ends are far from the image projection plane compared to $X_0=$ 0 m as in Fig. \ref{proj_2}. Although the distortion in ISAR images exists, it is quite minor, and the ISAR images have nearly the same scaling compared to the point scatterer models. Thus, it is clear that the proposed ISAR imaging technique works well with high resolution.\footnote{The additional simulation results in the supplementary document show that the proposed technique works well for other vehicular model as well.} Furthermore, Fig. \ref{proj_2} shows the dominant scatterers that are closer from the RSU are located in the lower part of image projection plane, which occurs the flipping of ISAR images.

\section{Conclusion}
In this paper, we developed the ISAR image formation technique based on the JRCS for vehicular environments exploiting the IEEE 802.11ad waveform. The range profile for multiple dominant scatterers of a vehicle was accurately reconstructed through the good correlation property of Golay complementary sequences and the extremely high carrier frequency of the IEEE 802.11ad waveform. The Doppler shifts were estimated using LSE for the echo signals of a specific frame with distinct delays based on the assumption of fixed vehicular velocity during the sufficiently short CPI of interest. Then, the phase wrapping was compensated using the Doppler shift estimates of properly separated frames. After recovering the vehicular velocity from the estimated Doppler shifts and equation of motion, the RSU could obtain the ISAR images from the cross-range FFT of pre-image matrix consisting of the range and cross-range profiles. Finally, we demonstrated the effectiveness of proposed ISAR imaging scheme via simulations with realistic vehicular environment models having short CPI. The proposed ISAR imaging can be exploited in many vehicular applications, e.g., traffic tendency mapping, vehicle type cognition, pedestrian safety system, and more, independent of weather conditions.

There are many possible future topics for the radar imaging based on the JRCS. First of all, the proposed technique can be easily adaptive to other communication waveform with some modifications. For example, more advanced WLAN standard 802.11ay has the preamble compatible to IEEE 802.11ad, and the channel bonding in 802.11ay can be exploited to form ISAR images with even higher resolution \cite{Ghasempour:2017}. The other examples are optimal waveform design for the JRCS ISAR imaging, the JRCS imaging using unmanned aerial vehicles (UAVs), exploiting future wireless communication features including intelligent reflecting surfaces (IRSs), and more.


\ifCLASSOPTIONcaptionsoff
\newpage
\fi

\bibliographystyle{IEEEtran}
\bibliography{refs_all}

\end{document}